\documentclass[12pt]{JHEP}
\usepackage{epsfig}

\title{Prototype Graphs for Radiative Corrections to 
Polarized Chargino or Neutralino Production in Electron-Positron
 Annihilation }
\author{M.A. Diaz \\ Departamento de Fisica, Pontificia Universidad
 Cat\'{o}lica de Chile,  Santiago 690441, Chile.}
\author{D.A. Ross \thanks{On leave of absence from:
 Department of Physics and Astronomy, University of Southampton,
 Southampton SO17 1BJ, U.K.}
 \\ Division Th\'{e}orique, CERN, 1211
 Geneva 23, Switzerland}

\abstract{We present the contributions from {\it all}
types of one-loop corrections to the scattering amplitude
for the pair production of polarized charginos or neutralinos
from polarized electron-positron annihilation. The contributions
are classified in terms of ``prototypes'' distinguished by the
number of particles inside the loops and their spins. The results
are quoted in terms of the Veltman-Passarino functions in terms of
general couplings and internal masses. The results can therefore
be applied to any supersymmetric extension of the Standard 
Model or indeed to any polarized fermion pair production process
in electron-positron annihilation.}
\preprint{CERN-TH/2001-071 \\
   UCCHEP/16-01}

\def\beq{\begin{equation}}
\def\eeq{\end{equation}}
\def\bea{\begin{eqnarray}}
\def\eea{\end{eqnarray}}

\def\ma{m_{\chi_a}}
\def\mb{m_{\chi_b}}
\def\sn{\sin\theta}
\def\cs{\cos\theta}

\begin{document}

\catcode`\@=11
\@addtoreset{equation}{section}
\@addtoreset{figure}{section}

\renewcommand{\theequation}{\thesection.\arabic{equation}}

\section{Introduction}
The results from the last days of the running of LEP, which suggest
that the mass of the Higgs scalar is around 115 GeV, has further
 enhanced the expectation that supersymmetry is realized in Nature
\cite{eno,kk}. A Supersymmetric extension of the Standard Model
is the only known scenario which can stabilise the effective potential
whilst maintaining the experimental limits on the gauge-boson masses
\cite{er}, as well as providing the possibility for
a scenario of  Grand Unification with a natural heirachy. 

Nevertheless, a confirmation of supersymmetry can only result from
the experimental identification of supersymmetric partners. The cleanest
signal comes from the fermionic superpartners of the gauge
 bosons and Higgs bosons, the charginos and neutralinos.
Regrettably LEP produced no evidence whatsoever for the existence
 of such particles and their existence is expected to be confirmed
or negated by the next generation of hadron colliders.
Notwithstanding this, the cleanest signal is still expected from
electron-positron colliders and as such a possible future linear collider
has an important role to play in the identification and study of
fermionic superpartners. 

The production cross-section for these superpartners is sensitive
to the parameters of the supersymmetry extension. Conversely, a study
of the production cross-section can be used to deduce information
about the supersymmetry parameters. As has been shown by Choi. et. al.
\cite{kal}, even more information can be gleaned from
 the study of cross-sections  in which both the initial
electron and positron beams are polarized {\it and} the spin
polarization of the final state fermion is also measured.  

The polarized cross-section was studied in detail at the tree-level in
\cite{kal}. It is known, however, \cite{dkr, dkr2} that at least for 
certain regions of parameter space, the higher order corrections
to the production cross-section can be substantial and they are also
sensitive to supersymmetry breaking parameters such as the
 squark masses, which do not enter into the tree-level calculation. 

Recently Blank and Hollick \cite{bh}
have reported on  a complete calculation of the 
higher order corrections to the
forward-backward asymmetry in the case of polarized electron and positron
beams, but did not consider the polarization of the final state fermions. 

Reports of numerical calculations for higher order corrections are
 undoubtedly very useful, but in the case of supersymmetric theories they
inevitably refer to a particular set of parameters and need to be
entirely repeated should one wish to calculate the effect of higher order
corrections in a different part of parameter space.

For this reason we present in this paper the contributions to the 
polarized scattering amplitudes from {\it all} the one-loop prototype
diagrams that can occur in a supersymmetric extension of
the Standard Model. A prototype diagram is defined by the spins of the
internal particles of the loop and we present the results in terms of
the Veltman-Passarino functions \cite{vp} for completely general
values of the couplings and masses of the internal particles.
The formulae presented in this paper can be applied 
to polarized chargino or neutralino production in electron-positron
annihilation, or indeed to any process in which a pair of
 massive fermions is produced and their polarization measured.  

Although the total correction to the cross-section is inevitably
gauge invariant, the contribution from individual graphs is not, 
and we work consistently in the `t Hooft-Feynman gauge, 
in which a gauge boson with momentum $k$ and mass $M$ has the propagator
$$ \frac{- \, i \, g^{\mu\nu}}{(k^2-M^2)}. $$ 
Some of the Veltman-Passarino functions are ultraviolet divergent.
It is assumed that these divergences have been {\it regulated} 
using dimensional regularization. On the other hand the renormalization
scheme is, in general,  unspecified and is reflected in the precise
value assigned to the finite parts of the ultraviolet divergent 
Passarino-Veltman functions. Mass renormalizations of
 self-energy insertions are physical (i.e. the renormalization
has been affected on mass-shell) and physical wavefunction
renormalization has been carried out for the external particles.

The mass of the electron is neglected throughout. 

The paper is organized as follows: In section 2 we
discuss the general formalism for the calculation
of the polarized scattering amplitudes in terms of a set of
 coefficient functions $ {\mathcal Q}_{L(R)i,j}$, which are a 
generalization of the coefficient functions used in \cite{kal}
for the tree-level calculation. These coefficient functions
multiply matrix elements of various Dirac-matrix structures,
 some of which have been relegated to Appendix A.
In section 3, we discuss the contribution to these coefficient functions
from self-energy and vertex-correction graphs in terms of the more 
familiar language of form-factors. The relations between the form-factors and the coefficient functions is given.
The detailed contributions from the various prototype graphs
to the self-energies are given in Appendix B, to the vertex corrections
in Appendix C, and to the box-diagrams, which must be expressed directly
in terms of the coefficient functions, in Appendix D.
In section 4 we present our conclusions.

\section{Generalized polarized scattering amplitudes}

In this section we restrict our discussion to charginos, although
it can be equally applied to neutralino production.

We write the scattering amplitude for an electron with
helicity $\alpha=R,L$ and momentum $p_1$,
 and a positron of opposite helicity and momentum $p_2$ into a
(positively charged) chargino of mass $\mb$ with momentum $k_2$
and helicity $\lambda_2$ and a (negatively charged) anti-chargino
of mass $\ma$, momentum $k_1$ and helicity $\lambda_1$ as 
\beq {\cal A}^\alpha_{\lambda_2,\lambda_1} \ = \ 
   \frac{2}{s} L^\mu_{\alpha}\, 
    Q_{\mu \, \alpha}^i \, \langle k_2, \lambda_2 | \Gamma^i
 | k_1, \lambda_1 \rangle . \label{me1}. \eeq
These amplitudes are normalized as in \cite{kal}, such that the differential
cross-section is given by
\beq \frac{d\sigma(\alpha,\lambda_2,\lambda_1)}{d\cos\theta}
  \ = \ \frac{\lambda^{1/2}(s,\ma^2,\mb^2)}{128 \, \pi \, s} 
 \left| {\cal A}^\alpha_{\lambda_2,\lambda_1} \right|^2, \eeq
where
$$ \lambda(x,y,z) \ \equiv \ x^2  +  y^2  +  z^2 
  -  2  x  y     -  2  x  z  -  2  y  z $$

The contribution from any Feynman graph to such an amplitude can always
be expressed in this form by making a suitable Fierz transformation where
necessary.
Here $ L^\mu_{R(L)}$ is the leptonic matrix element
$$  L^\mu_{R(L)} \ = \ \bar{v}(p_2) \gamma^\mu \frac{(1\pm \gamma^5)}{2}
    u(p_1). $$
Since the leptons are considered to be massless these two are the only
possible structures for the lepton factor. On the other hand the chargino
factor is the sum of matrix elements of five possible
 $\gamma-$matrix structures $\Gamma_i, \ i=1 \cdots 5$ are given by
\begin{eqnarray} \Gamma^1 & = & \frac{(1+\gamma^5)}{2} \nonumber \\
 \Gamma^2 & = & \frac{(1-\gamma^5)}{2}  \nonumber \\
 \Gamma^3 & = & \gamma^\nu\frac{(1+\gamma^5)}{2}  \nonumber \\
 \Gamma^4 & = & \gamma^\nu\frac{(1-\gamma^5)}{2}  \nonumber \\
 \Gamma^5 & = & -i \, \sigma^{\nu\rho} \end{eqnarray}
The coefficients $ Q_{\mu \, \alpha}^i$ are tensors which can be reduced
to the following structures, in terms of scalar quantities
${\cal Q}_{L(R)i,j}, \ j=1,2$,  as follows
\begin{eqnarray}
Q^\mu_{L(R)1} & = & {\mathcal Q}_{L(R)1} \, k_-^\mu \ 
  \nonumber \\
   Q^\mu_{L(R)2} & = & {\mathcal Q}_{L(R)2} \, k_-^\nu \ 
    \nonumber \\
Q^{\mu\nu}_{L(R)3} & = & {\mathcal Q}_{L(R)3,1} \, g^{\mu\nu} \  \, + 
             {\mathcal Q}_{L(R)3,2} \, k_-^\mu \,  p^\nu \nonumber \\
Q^{\mu\nu}_{L(R)4} & = & {\mathcal Q}_{L(R)4,1} \, g^{\mu\nu} \  \, + \ 
             {\mathcal Q}_{L(R)4,2} \, k_-^\mu \, p^\nu \nonumber \\
Q^{\mu\nu\rho}_{L(R)5} 
& = & {\mathcal Q}_{L(R)5,1} \, g^{\mu\nu} \, p^\rho \  \, -  \ 
   i \,  {\mathcal Q}_{L(R)5,2} \, \epsilon^{\mu\nu\rho\tau} \, p_\tau,
   \end{eqnarray}
where $ k_-^\mu=(k_1^\mu-k_2^\mu)$ and $ p^\mu=(p_1^\mu-p_2^\mu)$.
Any other structure can be expressed in terms of the above quantities,
by exploiting the fact that the leptonic current is conserved and that
the matrix elements of  $\Gamma^i$ are taken between on-shell chargino
states. We note here that at the tree level, only 
${\mathcal Q}_{L(R)3,1}$ and ${\mathcal Q}_{L(R)4,1}$ are non-zero.
Furthermore ${\mathcal Q}_{L(R)3,2}, \ {\mathcal Q}_{L(R)4,2}, \ 
{\mathcal Q}_{L(R)5,1}$   and ${\mathcal Q}_{L(R)5,2}$ do not occur
in self-energy or vertex correction graphs, but only arise
when boxes are taken into consideration.

We express the contributions to the scattering amplitude 
multiplying the scalar quantities ${\cal Q}_{L(R)i,j}$, using the following
notation: \\
$k$ is the 3-momentum of the $\chi$ in the  C.M. frame and
$\theta$ is the scattering angle in the C.M. frame, i.e. the angle
between the incident electron and the final-state $\chi^-$. 
We define further:
\beq  f_+ \ = \ \sqrt{(s-\ma^2-\mb^2)+2\, k \, \sqrt{s}} \eeq
\beq  f_- \ = \ \sqrt{(s-\ma^2-\mb^2)-2\, k \, \sqrt{s}} \eeq

The scattering amplitude contributions from each of the
${\cal Q}_{L(R)i,j}$ are  given by:
\beq - \, {\cal Q}_{L(R)1} 
 \left\{ \frac{2 \sqrt{2} \, k \, \sin\theta}{\sqrt{s}}
 \left( f_- \delta_{\lambda_2 +}\delta_{\lambda_1 +} \, + \,
 f_+ \delta_{\lambda_2 -}\delta_{\lambda_1 -} \right) \right\}
 \label{meq1}, \eeq
\beq  {\cal Q}_{L(R)2} \left\{ \frac{2 \sqrt{2} \, k \, \sin\theta}{\sqrt{s}}
 \left( f_+ \delta_{\lambda_2 +}\delta_{\lambda_1 +} \, + \,
 f_- \delta_{\lambda_2 -}\delta_{\lambda_1 -} \right) \right\}
 \label{meq2}, \eeq
\beq - \, {\cal Q}_{R,3,1} \left\{ \frac{2 \,  k \, \sin\theta}{\sqrt{s}}
 \left(a_1(R,\lambda_2,\lambda_1)-a_2(R,\lambda_2,\lambda_1)\right) 
 - \, 2 \, a_4((R,\lambda_2,\lambda_1) \right\} \label{meq3}, \eeq
\beq - \, {\cal Q}_{L,3,1} \left\{ \frac{2 \,  k \, \sin\theta}{\sqrt{s}}
 \left(a_1(R,\lambda_2,\lambda_1)-a_2(R,\lambda_2,\lambda_1)\right) 
 + \, 2 \, a_4((R,\lambda_2,\lambda_1) \right\} \label{meq4}, \eeq
\beq - \, {\cal Q}_{R(L),3,2}
   \left\{ \frac{4 \, k \, \sn}{\sqrt{s}} 
   \tilde{a}_3(R,\lambda_2,\lambda_1) \right\}, \label{meq5} \eeq
\beq - \, {\cal Q}_{R,4,1} \left\{ \frac{2 \,  k \, \sin\theta}{\sqrt{s}}
 \left(a_1(L,\lambda_2,\lambda_1)-a_2(L,\lambda_2,\lambda_1)\right) 
 - \, 2 \, a_4((L,\lambda_2,\lambda_1) \right\} \label{meq6}, \eeq
\beq - \, {\cal Q}_{L,4,1} \left\{ \frac{2 \,  k \, \sin\theta}{\sqrt{s}}
 \left(a_1(L,\lambda_2,\lambda_1)-a_2(L,\lambda_2,\lambda_1)\right) 
 + \, 2 \, a_4((L,\lambda_2,\lambda_1) \right\} \label{meq7}, \eeq
\beq - \, {\cal Q}_{R(L),4,2}
   \left\{ \frac{4 \, k \, \sn}{\sqrt{s}} 
  \tilde{a}_3(L,\lambda_2,\lambda_1) \right\}, \label{meq8} \eeq
\begin{eqnarray} & &   {\cal Q}_{R,5,1} \Bigg\{  \sqrt{2}
 \left( f_+-f_- \right) \Bigg[
 \sn \, \delta_{\lambda_2\lambda_1}
\nonumber \\ & & \hspace*{3cm}
 \,  + \,    \frac{(\ma-\mb)}{\sqrt{s}}
     \left( \cs \, \mathrm{ sgn}(\lambda_1) \, + \, 1 \right)
       \delta_{\lambda_2,-\lambda_1} \Bigg] 
\Bigg\}, \label{meq9} \end{eqnarray}
\begin{eqnarray}  & & - \,  {\cal Q}_{L,5,1} \Bigg\{  \sqrt{2}
 \left( f_+-f_- \right) \Bigg[ \sn \, \delta_{\lambda_2\lambda_1}
 \nonumber \\ & & \hspace*{3cm}
\,  + \,    \frac{(\ma-\mb)}{\sqrt{s}}
\left(   \cs \, \mathrm{ sgn}(\lambda_1) \, - \, 1 \right)
    \delta_{\lambda_2,-\lambda_1} \Bigg]
 \Bigg\} \label{meq10}, \end{eqnarray}
\begin{eqnarray}  & & - \, {\cal Q}_{R,5,2} \Bigg\{ 2 \, \sqrt{2}
 \left( f_++f_- \right) \Bigg[ \sn \,
    \delta_{\lambda_2\lambda_1} \mathrm{ sgn}(\lambda_1)
   \nonumber \\ & & \hspace*{3cm} -\frac{(\ma+\mb)}{\sqrt{s}} 
 \left(\cs \,  + \,  \mathrm{ sgn}(\lambda_1) \right)
  \delta_{\lambda_2,-\lambda_1}
   \Bigg] \Bigg\} \label{meq11}, \end{eqnarray}
\begin{eqnarray}  & & {\cal Q}_{L,5,2} \Bigg\{ 2 \, \sqrt{2}
 \left( f_++f_- \right) \Bigg[ \sn \,
    \delta_{\lambda_2\lambda_1} \mathrm{ sgn}(\lambda_1) 
   \nonumber \\ & & \hspace*{3cm} -\frac{(\ma+\mb)}{\sqrt{s}} 
 \left(\cs \, - \,  \mathrm{ sgn}(\lambda_1) \right)
  \delta_{\lambda_2,-\lambda_1}  
 \Bigg] \Bigg\} \label{meq12}. \end{eqnarray}
The amplitudes,  $ a_i(\alpha,\lambda_2,\lambda_1), 
 , \ (i=1\cdots 4)$
and $\tilde{a}_3((\alpha,\lambda_2,\lambda_1)$
 are given in  Appendix A.

\section{Self-energy and vertex corrections}
The coefficient functions,  ${\cal Q}_{L(R)i,j}$,
which acquire contributions from self-energy insertions and
triangle graphs are best treated in terms of form-factors, which replace
point-like vertices in the tree-level contributions that arise
either from a graph involving an exchange in the $s-$channel of 
a vector boson, or the $t-$channel exchange of a scalar particle
( a sneutrino in the case of chargino pair production from
electron-positron annihilation).

We write the general vertex for a vector boson coupling to 
an anti-fermion momentum fermion with (on-shell) 
momenta $k_1$ and $k_2$ respectively, as
$$ F_0^+ \,\gamma^\mu \frac{(1+\gamma^5)}{2}
\, + \,   F_0^- \,\gamma^\mu \frac{(1-\gamma^5)}{2}
\, + \, F_k^+ \,k_-^\mu \frac{(1+\gamma^5)}{2}
\, +  \, F_k^- \,k_-^\mu \frac{(1-\gamma^5)}{2}. $$ 
Note that the form-factors $ F_k^+,  F_k^-$ are absent at tree-level.

The Yukawa coupling of
fermions to a scalar particle is given by
$$ F_{\tilde{\nu}}^\pm \frac{(1\pm \gamma^5)}{2} $$
for right-(left-) handed incoming fermions respectively.
 
Thus the contributions to  the coefficient functions from the exchange
of a vector boson of mass $M_V$ in the $s-$channel are:
\begin{eqnarray}
 \Delta  {\cal Q}_{L(R)3,1} & = & g_{L(R)} \frac{s}{(s-M_V^2)}
  F_{0 \, (V)}^+, \nonumber \\
 \Delta  {\cal Q}_{L(R)4,1} & = & g_{L(R)} \frac{s}{(s-M_V^2)}
  F_{0 \, (V)}^-,  \nonumber \\
 \Delta  {\cal Q}_{L(R)1} & = & g_{L(R)} \frac{s}{(s-M_V^2)}
  F_{k \, (V)} ^+, \nonumber \\
 \Delta  {\cal Q}_{L(R)2} & = & g_{L(R)} \frac{s}{(s-M_V^2)}
  F_{k \, (V)}^-, \end{eqnarray}
where $g_{L(R)}$ is the coupling of the left-(right-)handed electrons
to the vector boson.

Upon performing a Fierz transformation, the contributions to the
 coefficient functions from the exchange of a scalar particle
of mass $m_{\tilde{\nu}}$ in the $t-$channel, is given by
\beq \Delta  {\cal Q}_{L \, 4,1} \ = \ - \frac{1}{2} \, 
 \frac{s}{(t-m_{\tilde{\nu}}^2)}
   F_{\tilde{\nu}}^- ( F_{\tilde{\nu}}^-)^\dagger, \eeq
for a scalar particle that couples to left-handed incoming electrons
and
\beq \Delta  {\cal Q}_{R \, 3,1} \ = \  - \frac{1}{2} \, 
 \frac{s}{(t-m_{\tilde{\nu}}^2)}
   F_{\tilde{\nu}}^+ ( F_{\tilde{\nu}}^+)^\dagger, \eeq
for a scalar particle that couples to right-handed incoming electrons.

Note that $F^\pm_{\tilde{\nu}}$ are the form-factors for an incoming electron
and outgoing anti-chargino or anti-neutralino with helicities $\pm$,
whereas $(F^\pm_{\tilde{\nu}})^\dagger$
 are the form-factors for an incoming positron
and outgoing chargino or neutralino with helicities $\pm$,

\subsection{Self-energy insertions:}

The contribution to the form-factors $F_{\tilde{\nu}}^\pm$
from the self-energy $\Sigma_{\tilde{\nu}}(t)$ of the scalar
particle exchanged in the $t-$channel is given by

\beq \Delta F_{\tilde{\nu}}^\pm \ = 
\ \frac{\left(\Sigma_{\tilde{\nu}}(t)-\Sigma_{\tilde{\nu}}
(m_{\tilde{\nu}}^2)
\right)}{\left(t-m_{\tilde{\nu}}^2 \right)} \, F_{\tilde{\nu}}^\pm. \eeq

For the exchange of a gauge-boson in the $s$-channel, we need to account
 for possible mixing at the one-loop level between a gauge-boson $V$
and a gauge-boson $V^\prime$. The transverse part of the
self-energy can be written in general
as $\Sigma_{VV^\prime}(s)$, and the contribution to the form-factors
$F_{0 \, (V)}^\pm$ is given by
 
\beq \Delta F_{0 \, (V)}^\pm \ = 
\ \frac{\left(\Sigma_{VV^\prime}(s)-\Sigma_{VV^\prime}
(M_{V^\prime}^2)
\right)}{\left(s- M_{V^\prime}^2 \right)} \,  F_{0 \, (V^\prime)}^\pm. 
\eeq

The contribution to the form-factors from  the self-energy of 
the charginos (or neutralinos) is more complicated owing to 
different one-loop contributions to left-and right-handed components
and  to the mixing between the different species of charginos 
(neutralinos). The was discussed in detail in \cite{dkr}. Here we
reproduce the main results.

The general self-energy of a chargino or neutralino of momentum $p$
can be written as
\beq  \Sigma_{ab}(p) \ = \ 
 \left( A_{ab}^+(p^2) +
 B_{ab}^+(p^2) \, \gamma \cdot p \right)
              \frac{(1+\gamma^5)}{2}
 \, + \,  \left( A_{ab}^-(p^2) +
    B_{ab}^-(p^2)  \, \gamma \cdot p \right) \frac{(1-\gamma^5)}{2}, \eeq
where the suffix $ab$ refers to an outgoing chargino (neutralino) of type
$a$ and an incoming chargino  (neutralino) of type $b$. 

We denote $F^{\pm \, a }$ to mean the form-factor involving a chargino
or neutralino of type $a$ (with mass $m_a$).

The correction to the form-factors due to the self-energies
of the external fermion is given by

\begin{eqnarray} \Delta F^{\pm \, a } &=&
\frac{1}{2}\left\{ 
 \pm \frac{(A^+_{aa}-A^-_{aa})}{2 m_a} \, + 
 B^\pm_{aa} \, + \, m_a \left( A^{+ \, \prime}_{aa} +
 A^{- \, \prime}_{aa} + m_a (
  B^{+ \, \prime}_{aa} +
 B^{- \, \prime}_{aa} ) \right) \right\} F^{\pm \, a }
 \nonumber \\
 & +  &  \sum_{b\neq a}
\left\{ \frac{   m_a A^\pm_{ab} +  m_b A^\mp_{ab}
+ m_a^2 B_{ab}^\pm + m_a m_b B_{ab}^\mp
   }{(m_a^2-m_b^2) }  \right\}  F^{\pm \, b } \end{eqnarray}
for the self-energy correction to a chargino (neutralino), and
\begin{eqnarray} \Delta F^{\pm \, a } &=&
\frac{1}{2}\left\{ 
 \pm \frac{(A^-_{aa}-A^+_{aa})}{2 m_a} \, + 
 B^\pm_{aa} \, + \, m_a \left( A^{+ \, \prime}_{aa} +
 A^{- \, \prime}_{aa} + m_a (
  B^{+ \, \prime}_{aa} +
 B^{- \, \prime}_{aa} ) \right) \right\}  F^{\pm \, a } 
 \nonumber \\
 & +  &  \sum_{b\neq a}
\left\{ \frac{  m_a A^\mp_{ba} +  m_b A^\pm_{ba}
+ m_a^2 B^\pm_{ba} + m_a m_b B^\mp_{ba}
   }{(m_a^2-m_b^2) }  \right\}  F^{\pm \, b } 
 \end{eqnarray}
for the self-energy correction to an anti-chargino (anti-neutralino).
The $\prime$ refers to the derivative w.r.t. $p^2$
and all the functions are taken at the point $p^2=m_a^2$.

The contributions from the various prototype diagrams to the self
 energies, $\Sigma_{\tilde{\nu}}, \ \Sigma_{VV^\prime}, \  \Sigma_{ab}$  
are given in Appendix B.

\subsection{Triangle graphs:}
The contributions to the form-factors, 
$ F_0^\pm, \   \  F_k^\pm, \   F_{\tilde{\nu}}^\pm$ from the different
possible triangle graphs are given in Appendix C.   

We note here that for a complete one-loop calcualtion, we also
need to account for the form-factor of the vertex between the intermediate
gauge bosons and the incoming electron-positron pair. Since we neglect the
 masses of the leptons only the form-factors, $F_0^\pm$ are non-zero.
The contributions to these are obtained form the vertex protoypes 5 - 8
by setting $\ma, \, \mb \, = \, 0$.

\section{Box graphs}
The contibutions to the coefficient functions  $ {\mathcal Q}_{L(R)i,j}$
from all the prototype box graphs are given in Appendix D. All box graphs
contributing to the process under consideration can be expressed in terms
of these prototypes or prototypes which can be obtained from them by
the simple crossing relations $t \, \leftrightarrow \, u$,
 $\ma \, \leftrightarrow \, \mb$. 

\section{Conclusions}
In this paper we have presented explicit expressions for the contributions
from {\it all} prototypes of one-loop corrections in terms of the
Veltman-Passarino functions. The results are given in terms of general
couplings and internal masses, so that they can be used in any
model for the determination of the scattering amplitudes of
polarized fermion pair production.

The results are presented in two stages. Firstly, the helicity amplitudes
 are given in terms of a small number of coefficient functions, 
 $ {\mathcal Q}_{L(R)i,j}$, of various effective interaction structures.
 Secondly the contributions from all prototype Feynman diagrams to these
coefficient functions are given.

There exist \cite{vp} relations between the Veltman-Passarino functions 
used in this paper, such that they can all be reduced
 to ``scalar'' types functions, $A0, \ B0, C0 $ and $D0$.
 We have chosen {\it not} to implement these relations in the
 expressions presented here. With the exception of cases in which some
 of the internal particles masses vanish, the expressions tend be be 
considerably longer when only this minimal set of functions is used.
Instead, it is probably more efficient to implement these relations
at the point of numerical calculations. A FORTRAN library exists
\cite{ff} for the numerical calculation of the Veltman-Passarino
 functions, but unfortunately these only calculate such functions
with up to one power of the loop momentum in the numerator.
We have written a set of FORTRAN subroutines which are to be used
 in conjunction with this library and can be used to calculate
the Veltman-Passarino functions with up to three powers of loop momenta
in the numerator (for the calculation addressed in this paper only
two powers of loop momenta are required). These routines
are available upon request as well as a set of subroutines that
calculate the contributions to the coefficient functions,
$ {\mathcal Q}_{L(R)i,j}$,
from all the prototype diagrams considered in this paper.

\acknowledgments{
The authors wish to thank  Steve King for useful
conversations, assistance, and many helpful suggestions. \\
Many of the algebraic manipulations performed in this work were
carried out using FORM \cite{form}.} \\
This work was partially supported by CONICYT grant No: 1010974.

\section*{Appendix A}

\setcounter{equation}{0}
\renewcommand{\theequation}{A.\arabic{equation}}

The quantities $ a_i(\alpha,\lambda_2,\lambda_1)$ used in 
eqs.(\ref{meq3}-\ref{meq8})
are given by

\beq a_i(\alpha,\lambda_2,\lambda_1) \
 = R_i^j \,  \tilde{a}_j(\alpha,\lambda_2,\lambda_1) .\eeq

The $4 \times 4 $ matrix, $R$, can be written in terms
 of the  is the  the matrix  $3 \times 3 $ matrix, $T$ as
\beq R \ = \ \frac{1}{sk^2\sin^2\theta} \, 
\left( \begin{array}{cc}T & \\ & -k^2  \sin^2\theta
\end{array} \right), \eeq
where
\beq T= \left( \begin{array}{ccc}  
-\mb^2-k^2\cos^2\theta& \,
\frac{1}{2}s_{--}-k^2\cos^2\theta& \,
 \frac{k\cs}{2\sqrt{s}}s_{-+} \\
\frac{1}{2}s_{--}-k^2\cos^2\theta& \,
-\ma^2-k^2\cos^2\theta& \,
-  \frac{k\cs}{2\sqrt{s}}s_{+-} \\
 \frac{k\cs}{2\sqrt{s}}s_{-+} & \,
- \frac{k\cs}{2\sqrt{s}}s_{+-}& \,
-k^2  \end{array} \right), \eeq
with $s_{--}=s-\ma^2-\mb^2$, $s_{+-}=s+\ma^2-\mb^2$,
and $s_{-+}=s-\ma^2+\mb^2$. The quantities
  $ \tilde{a}_i(\alpha,\lambda_2,\lambda_1)$ are given 
(in four vector notation, $i=1 \cdots 4$) by
\begin{eqnarray}
 \tilde{a}(R,+,+) & = & \frac{1}{\sqrt{2}}\left(
  \ma f_+, \ \mb f_-, \ \left(\ma f_++\mb f_-\right)\cs,\  0 \right) 
 \nonumber \\ 
  \tilde{a}(R,+,-) & = & \frac{1}{\sqrt{2}}\left(
  0, \ 0, \ \sqrt{s}f_+\sn , \ -\sqrt{s}f_+   \right)  \nonumber \\ 
 \tilde{a}(R,-,+) & = & \frac{1}{\sqrt{2}}\left(
  0, \ 0, \ \sqrt{s}f_-\sn , \ \sqrt{s}f_-   \right)  \nonumber \\ 
 \ \tilde{a}(R,-,-) & = & \frac{1}{\sqrt{2}}\left(
  \ma f_-, \ \mb f_+, \ -\left(\ma f_-+\mb f_+\right)\cs,\  0 \right)
  \nonumber \\ 
 \tilde{a}(L,+,+) & = & \frac{1}{\sqrt{2}}\left(
- \ma f_-, \ -\mb f_+, \ \left(\ma f_-+\mb f_+ \right)\cs,\  0 \right)
  \nonumber \\ 
 \tilde{a}(L,+,-) & = & \frac{1}{\sqrt{2}}\left(
  0, \ 0, \ \sqrt{s}f_-\sn , \ -\sqrt{s}f_-   \right)  \nonumber \\ 
  \tilde{a}(L,-,+) & = & \frac{1}{\sqrt{2}}\left(
  0, \ 0, \ \sqrt{s}f_+\sn , \ \sqrt{s}f_+   \right)   \nonumber \\ 
 \tilde{a}(L,-,-) & = & \frac{1}{\sqrt{2}}\left(
  -\ma f_+, \ -\mb f_-, \ -\left(\ma f_++\mb f_-\right)\cs,\  0 \right)
 \end{eqnarray}
Note that the quantities $\tilde{a}_3(R(L)(\lambda_2,\lambda_1)$
of eqs. (\ref{meq5}, \ref{meq8}) are the third components of these vectors.

\section*{Appendix B}
{\bf Prototype self-energy  graphs} \bigskip

\setcounter{equation}{0}
\renewcommand{\theequation}{B.\arabic{equation}}

The self-energy corrections in a general supersymmetric theory have been
considered in detail in \cite{pb}. For completeness, and to conform with
the notation used in this paper, we repeat the results for the various 
self-energy prototype diagrams in this Appendix, in terms of the
Veltman-Passarino functions, $A0, \ B0, \  B1, \  B21, \ B22$, defined as

\begin{eqnarray} - \, i \,  \int \frac{d^4l}{(2\pi)^4} \frac{1}
{(l^2-m_1^2)((l+v)^2-m_2^2)  } & = &   
 \, B0  \nonumber \\
   - \, i \, \int \frac{d^4l}{(2\pi)^4} \frac{l^\rho}
{(l^2-m_1^2)((l+v)^2-m_2^2)  }  & = &   
   B1 v^\rho , \nonumber \\
  - \, i \, \int \frac{d^4l}{(2\pi)^4} \frac{l^2}
{(l^2-m_1^2)((l+v)^2-m_2^2)  }  & = &  
 A0(m_2^2)+m_1^2  \, B0  
\nonumber \\  - \, i \, \int \frac{d^4l}{(2\pi)^4} \frac{l^\rho l^\sigma}
{(l^2-m_1^2)((l+v)^2-m_2^2)  }  & = &  
 g^{\rho\sigma} B22 \, + \, B21 v^\rho v^\sigma.  \end{eqnarray}
\vspace*{1.5cm}

\subsection*{Scalar self-energy}

\vspace*{1cm}

{\bf SELF-ENERGY PROTOTYPE 1}

\FIGURE[h]{
\centerline{\epsfig{file=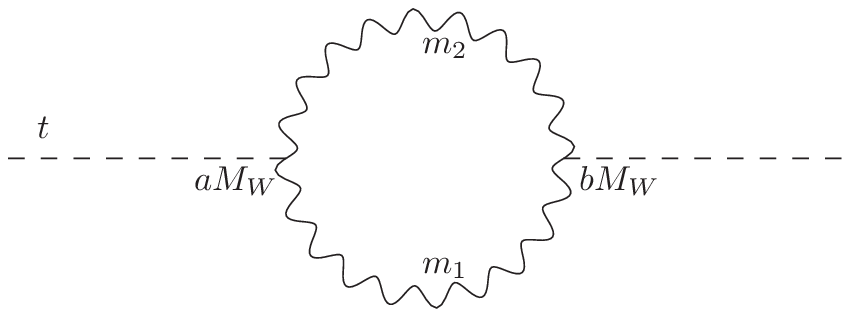}}
\caption{Self-energy prototype graph 1.}
\label{fig1}
}

\beq 16 \pi^2 \,  \Delta \Sigma_{\tilde{\nu}}(t) \ = \ 
 4 \, a \, b  \, M_W^2 \left[ B0 \, - \, \frac{1}{2} \right] \eeq
\newpage

{\bf SELF-ENERGY PROTOTYPE 2}

\FIGURE[h]{
\centerline{\epsfig{file=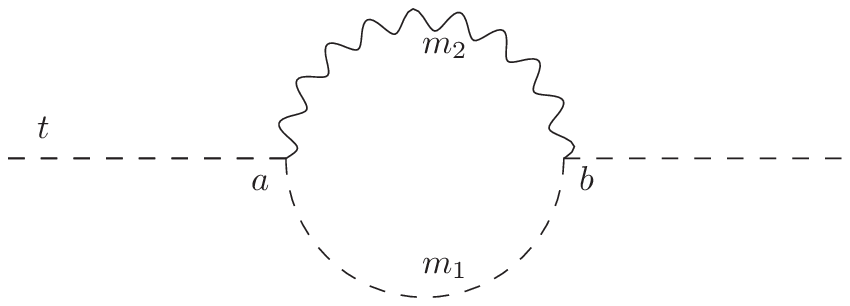}}
\caption{Self-energy prototype graph 2.}
\label{fig2}
}

\beq  16 \pi^2 \,  \Delta \Sigma_{\tilde{\nu}}(t) \ = \ 
 - \, a \, b  \, \left[ \left(t \, + \, m_1^2 \right) B0 
 \, - \, 2 \, t \, B1 + A0 \right] \eeq
\bigskip

{\bf SELF-ENERGY PROTOTYPE 3}

\FIGURE[h]{
\centerline{\epsfig{file=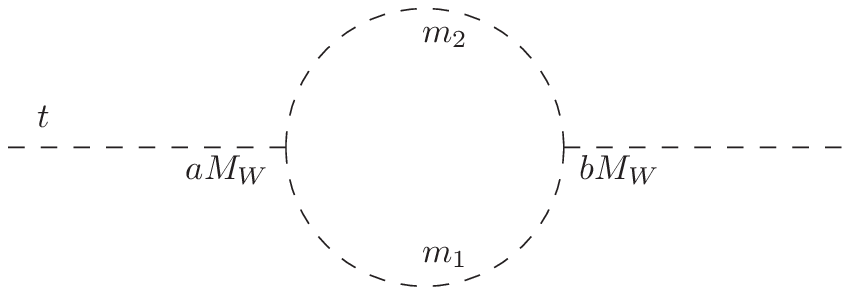}}
\caption{Self-energy prototype graph 3.}
\label{fig3}
}

\beq 16 \pi^2 \,  \Delta \Sigma_{\tilde{\nu}}(t) \ = \ 
  \, a \, b  \, M_W^2  B0  \eeq
\bigskip

{\bf SELF-ENERGY PROTOTYPE 4}

\FIGURE[h]{
\centerline{\epsfig{file=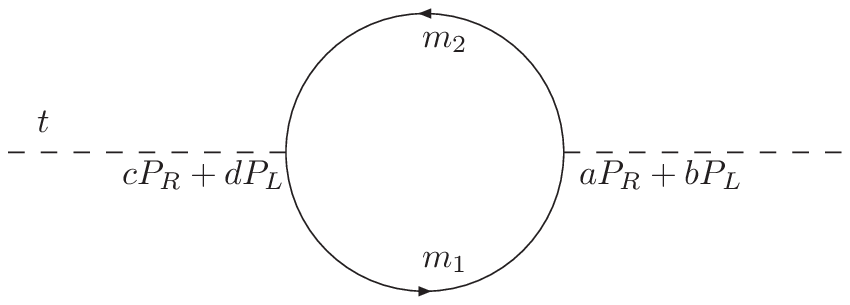}}
\caption{Self-energy prototype graph 4.}
\label{fig4}
}

\beq 16 \pi^2 \,  \Delta \Sigma_{\tilde{\nu}}(t) \ = \ 
 - \, 2  \left( a \, d  \, + \, b \, c  \right)
    \left[ A0 \, + \, m_1^2 B_0 \, + t \, B1 \right]
 \, - \, 2  \left( a \, c \, + \, b \, d  \right)
    m_1 m_2 B0  \eeq
\bigskip

\subsection*{Vector self-energy}

\vspace*{.6cm}

{\bf SELF-ENERGY PROTOTYPE 5}

\FIGURE[h]{
\centerline{\epsfig{file=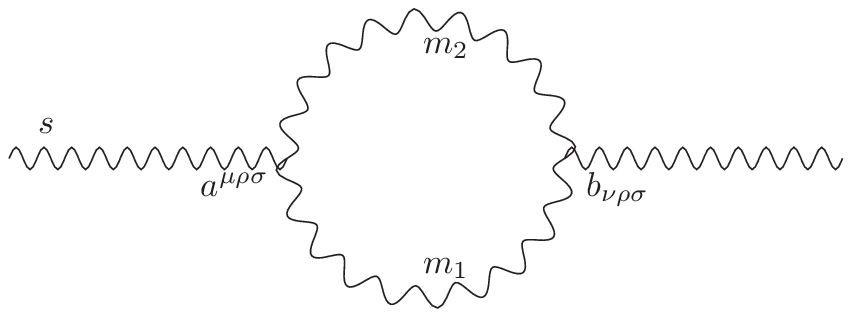}}
\caption{Self-energy prototype graph 5.}
\label{fig5}
}

$a^{\mu\rho\sigma}$ and $b_{\nu\rho\sigma}$ are the usual triple
gauge boson vertices with gauge couplings $a$ and $b$ respectively.
The contribution from Faddeev-Popov ghosts has been added to this graph.

\beq 16 \pi^2 \,  \Delta \Sigma_{VV\prime}(s) \ = \ 
 - \, a \, b  \left[ 8 \, B22 \, + \, 2 \, A0 \, 
+ \left( 5 \, s \, + 2 \, m_1^2  \right) B0 \, + 2 \, s \, B1 \, + \,
 \frac{2}{3}\, s \right] \eeq
\bigskip

{\bf SELF-ENERGY PROTOTYPE 6}

\FIGURE[h]{
\centerline{\epsfig{file=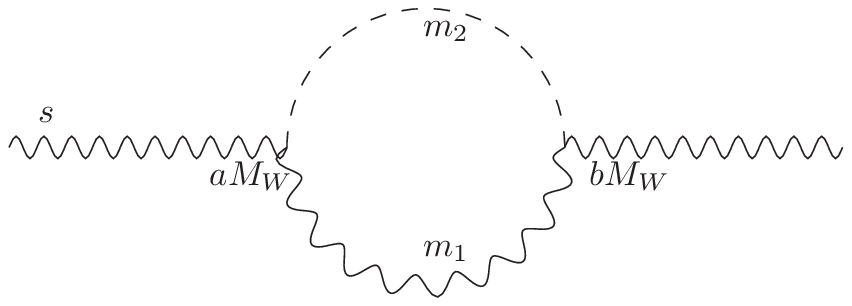}}
\caption{Self-energy prototype graph 6.}
\label{fig6}
}

\beq 16 \pi^2 \,  \Delta \Sigma_{VV\prime}(s) \ = \ 
  \, a \, b  \, M_W^2 B0 \eeq
\bigskip

{\bf SELF-ENERGY PROTOTYPE 7}

\FIGURE[h]{
\centerline{\epsfig{file=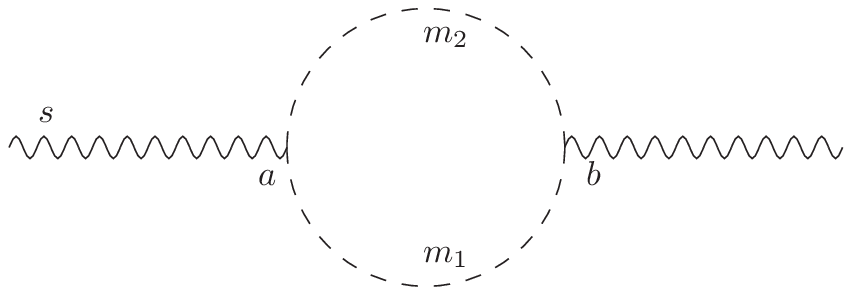}}
\caption{Self-energy prototype graph 7.}
\label{fig7}
}

\beq 16 \pi^2 \,  \Delta \Sigma_{VV\prime}(s) \ = \ 
 - \, 4  \, a \, b \,   B22 \eeq
\newpage

{\bf SELF-ENERGY PROTOTYPE 8}

\FIGURE[h]{
\centerline{\epsfig{file=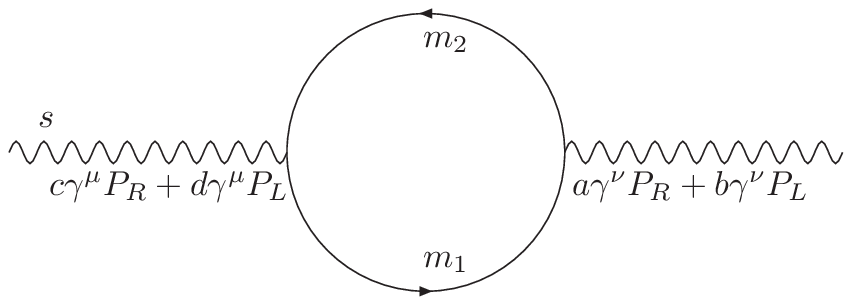}}
\caption{Self-energy prototype graph 8.}
\label{fig8}
}

\beq 16 \pi^2 \,  \Delta \Sigma_{VV\prime}(s) \ = \ 
 2 \left(a \, d \, + b \, c \right) m_1 m_2 B0 \, - \, 
2 \, \left(a \, c \, + \, b \, d \right) \left[   
    A0 \, + \, m_1^2 B0 \, + s \, B1 \, - \, 2 \, B22 \right]
 \eeq
\bigskip

\subsection*{Fermion self-energy}

\vspace*{1cm}

{\bf SELF-ENERGY PROTOTYPE 9}

\FIGURE[h]{
\centerline{\epsfig{file=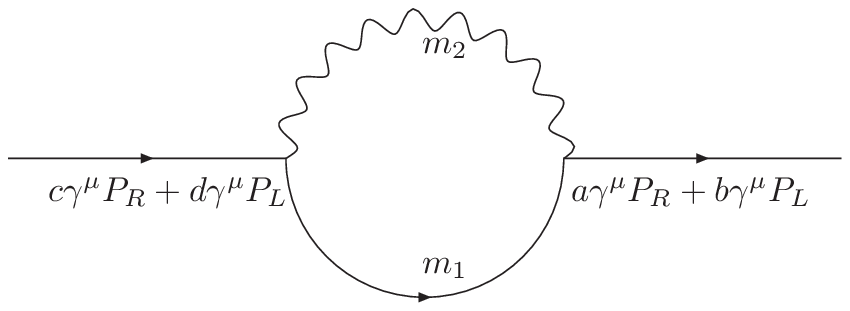}}
\caption{Self-energy prototype graph 9.}
\label{fig9}
}

\begin{eqnarray}
 16 \pi^2 \,  \Delta A^+ & = & 
 - \, 4 \, b \, c  \, m_1 \left[B0 \, - \, \frac{1}{2} \right] \nonumber \\
 16 \pi^2 \,  \Delta A^- & = & 
 - \, 4 \, a \, d  \, m_1 \left[B0 \, - \, \frac{1}{2} \right] \nonumber \\
 16 \pi^2 \,  \Delta B^+ & = & 
 - \, 2 \, a \, c   \left[B1 \, + \, \frac{1}{2} \right] \nonumber \\
 16 \pi^2 \,  \Delta B^- & = & 
 - \, 2 \, b \, d   \left[B1 \, + \, \frac{1}{2} \right] \end{eqnarray}
\bigskip

{\bf SELF-ENERGY PROTOTYPE 10}

\FIGURE[h]{
\centerline{\epsfig{file=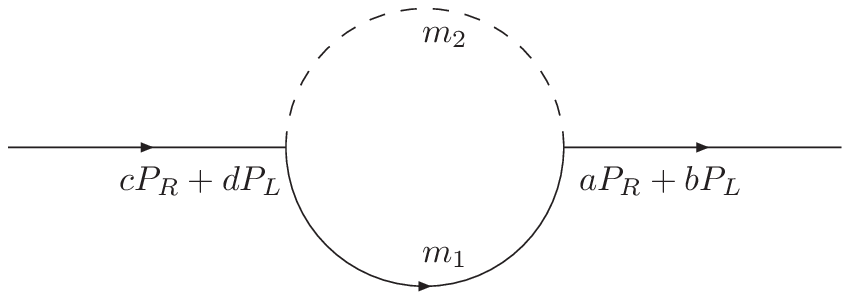}}
\caption{Self-energy prototype graph 10.}
\label{fig10}
}

\begin{eqnarray} 16 \pi^2 \,  \Delta A^+ & = & 
 a \, c  \, m_1 B0  \nonumber \\
 16 \pi^2 \,  \Delta A^+ & = & 
  b \, d  \, m_1 B0  \nonumber \\
 16 \pi^2 \,  \Delta B^+ & = & 
 - \,   b \, c  B1  \nonumber \\
 16 \pi^2 \,  \Delta B^- & = & 
 -  \, a \, d  \,  B1  \end{eqnarray}

\section*{Appendix C}
{\bf Prototype triangle  graphs} \bigskip

\setcounter{equation}{0}
\renewcommand{\theequation}{C.\arabic{equation}}

In this Appendix we give the expressions for the prototype
triangle diagrams, with couplings indicated in the diagrams, in terms
of the Veltman-Passarino functions, $C0, \ C11, \  C12, C21,$  $C23,
 \  C24$, These functions, with arguments 
$v_1^2, v_2^2, (v_1+v_2)^2, m_1^2, m_2^2, m_3^2$
are defined as
\begin{eqnarray}
 - \, i \,  \int \frac{d^4l}{(2\pi)^4} \frac{1}
{(l^2-m_1^2)((l+v_1)^2-m_2^2) \, (l+v_1+v_2)^2-m_3^2) } & = &   
 \, C0  \nonumber \\
   - \, i \, \int \frac{d^4l}{(2\pi)^4} \frac{l^\rho}
{(l^2-m_1^2)((l+v_1)^2-m_2^2) \, (l+v_1+v_2)^2-m_3^2) }  & = &   
   C11 \, v_1^\rho \, + \,   C12 \, v_2^\rho    , \nonumber \\
  - \, i \, \int \frac{d^4l}{(2\pi)^4} \frac{l^2}
{(l^2-m_1^2)((l+v_1)^2-m_2^2) \, (l+v_1+v_2)^2-m_3^2) }  & = &  
 B0+m_1^2  \, C0  \nonumber \\
    - \, i \, \int \frac{d^4l}{(2\pi)^4} \frac{l^\rho \, l^\sigma}
{(l^2-m_1^2)((l+v_1)^2-m_2^2) \, (l+v_1+v_2)^2-m_3^2) }  &= &   
 g^{\rho\sigma}  \, C24 \, +
  \, C21 v_1^\rho v_1^\sigma  \nonumber \\ & & \hspace*{-3cm}  +
  \, C22 v_2^\rho v_2^\sigma  \, + \, 
  \, C23\left( v_1^\rho v_2^\sigma +  v_2^\rho v_1^\sigma \right) 
    \end{eqnarray}

The arguments of the function $B0$ are $(v_2^2,m_2^2, m_3^2)$. 

For the form-factor of the scalar particles, exchanged in the $t-$channel,
one substitutes, $0, \ \ma^2$, (or $\mb^2$),  $t$, for
$v_1^2, \ v_2^2, \ (v_1+v_2)^2$ respectively, whereas for the 
vector-boson form factors one substitutes
$\ma^2, \  \mb^2, \  s$ for $v_1^2, \ v_2^2, \ (v_1+v_2)^2$ respectively.

The contributions  to the form-factors  from each prototype
vertex diagram are given below. Diagrams involving internal gauge-bosons
are calculated in Feynman gauge.

\vspace*{1.5cm}

\subsection*{Scalar form-factor diagrams}
\bigskip

{\bf VERTEX PROTOTYPE 1}

\FIGURE[h]{
\centerline{\epsfig{file=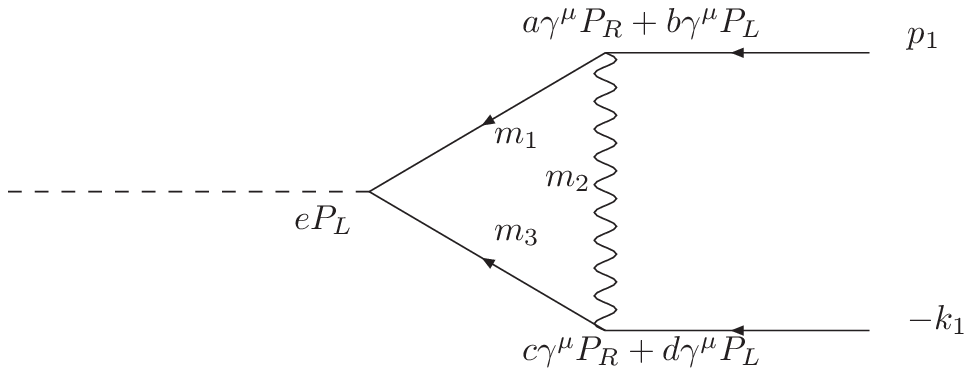}}
\caption{Vertex prototype graph 1.}
\label{figv1}
}

\beq 16\pi^2 \, \Delta F_{\tilde{\nu}}^- \  = \  
  \,   b \, d \, e \left[2   \, - \, 4 \, B0 \,
 + \, 2 \, \left( \ma^2-t \right) \left(C11 \, + \, C12 \right)
 \, - \, 4 \, \ma^2 C12 \right] \, + 2 \, 
\, b \, c \,  e \, m_3 \ma C12 \eeq
\bigskip

{\bf VERTEX PROTOTYPE 2}

\FIGURE[h]{
\centerline{\epsfig{file=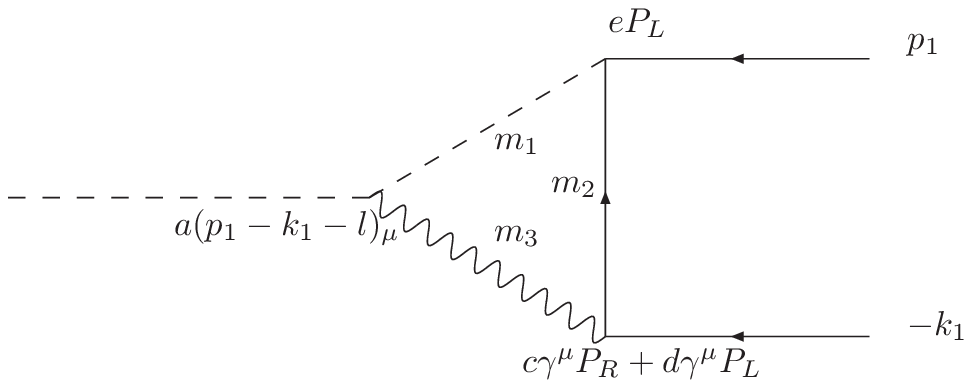}}
\caption{Vertex prototype graph 2.}
\label{figv2}
}

\beq 16\pi^2 \, \Delta F_{\tilde{\nu}}^- \  = \  
  \,   a \, d \, e \left[  \, B0 \,
 + \,  m_1^2 C0 \, -  \, t  \, C12 \right]
 \, + \, a \, c \,  e \, m_2 \ma \left[C0 \, - \,  C12 \right] \eeq
\bigskip

{\bf VERTEX PROTOTYPE 3}

\FIGURE[h]{
\centerline{\epsfig{file=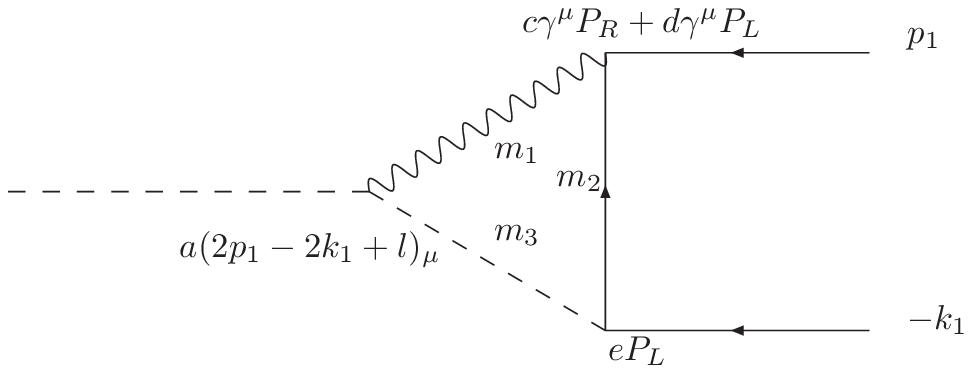}}
\caption{Vertex prototype graph 3.}
\label{figv3}
}

\beq 16\pi^2 \, \Delta F_{\tilde{\nu}}^- \  = \  
  \,   a \, d \, e \left[  \, B0 \,
 + \,  m_1^2 C0 \, + \, 2  \, \ma^2   \, C12 
 \, + \, \left(t-\ma^2\right) \left(C12 \, + 2 \, C11  \, - 
2 \, C0 \right)  \right] \eeq
\bigskip

{\bf VERTEX PROTOTYPE 4}

\FIGURE[h]{
\centerline{\epsfig{file=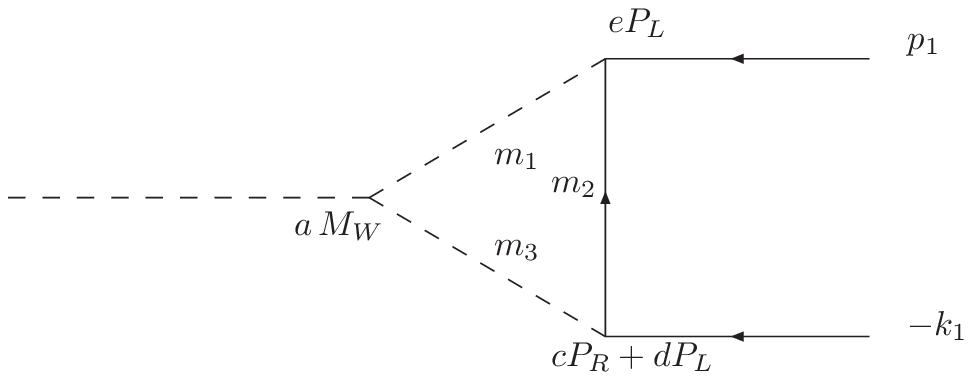}}
\caption{Vertex prototype graph 4.}
\label{figv4}
}

\beq 16\pi^2 \, \Delta F_{\tilde{\nu}}^- \  = \  
 2  \,   a \, d \, e \, M_W  m_2 C0 \, -  \,  2 \,
  a \, c \, e \, M_W \ma C12 \eeq
\bigskip

For all the above prototype graphs, we obtain
  similar expressions for $\Delta (F_{\tilde{\nu}}^-)^\dagger$
 with $\ma \to \mb$.
\bigskip

\subsection*{Vector form-factor diagrams}
\bigskip

{\bf VERTEX PROTOTYPE 5}

\FIGURE[h]{
\centerline{\epsfig{file=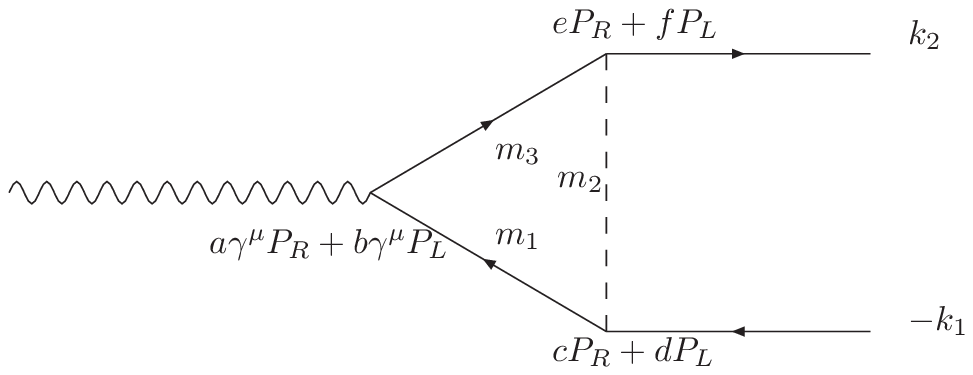}}
\caption{Vertex prototype graph 5.}
\label{figv5}
}

\begin{eqnarray} 16\pi^2 \, \Delta F_0^+ & = &  
    a \, c \, e \, \mb  m_3 C12 \, + \,  
 b\, d \, f \,  \ma m_3  C11  \nonumber \\ 
&+& a \, c \, f \left[ 2 \, C24 \, - \ \frac{1}{2}
 \, + \, \ma^2 \left( C11 \, - \, C12 \, + \, C21 \, - \, C23 \right)
 \right. \nonumber \\ & & \left. \hspace*{1cm}
 \, + \, \mb^2 \left( C22 \, - \, C23 \right) \, + \,  
s \left(C12 \, + C23 \right) \right] \nonumber \\
&+& b \, d \, e \, \ma \mb \left[ C11 \, -  \, C12 \right]
\, - \, b \, c \, f \, m_1 m_3 C0 \nonumber \\ &-&
a \, d \, f \, m_1 \ma \left[ C0 \, + \, C11 \right]
 \, - \, b \, c \, e  \,m_1 \mb \left[ C0 \, + C12 \right]
 \end{eqnarray}

\begin{eqnarray} 16\pi^2 \, \Delta F_0^- & = &  
    a \, c \, e \, \ma  m_3 C11 \, +  \,  
 b\, d \, f \,  \mb m_3  C12  \nonumber \\ 
&+& b \, d \, e \left[ 2 \,C24 \, - \ \frac{1}{2}
 \, + \, \ma^2 \left( C11 \, - \, C12 \, + \, C21 \, - \, C23 \right)
 \right. \nonumber \\ & & \left. \hspace*{1cm}
 \, + \, \mb^2 \left( C22 \, - \, C23 \right) \, + \,  
s \left(C12 \, + C23 \right) \right] \nonumber \\
&+& a \, c \, f \, \ma \mb \left[ C11 \, -  \, C12 \right]
\, - \, a \, d \, e \, m_1 m_3 C0 \nonumber \\ &-&
b \, c \, e \, m_1 \ma \left[ C0 \, + \, C11 \right]
 \, - \, a \, d \, f  \,m_1 \mb \left[ C0 \, + C12 \right]
 \end{eqnarray}

\begin{eqnarray} 16\pi^2 \, \Delta F_k^+ & = &  
a \, c \, f \, \mb \left[ C22 \, - \, C23 \right]
\, + \, b \, d \, e \, \ma \left[ C11 \, - \, C12 + \, C21 \, - \, C23
\right] \nonumber \\ &+& a \, c \, e \, m_3 C12 \, - \, 
 b \, c \, e \, m_1 \left[ C0 \, + \, C11 \right]  \end{eqnarray}

\begin{eqnarray} 16\pi^2 \, \Delta F_k^- & = &  
b \, d \, e \, \mb \left[ C22 \, - \, C23 \right]
\, + \, a \, c \, f \, \ma \left[ C11 \, - \, C12 + \, C21 \, - \, C23
\right] \nonumber \\ &+& b \, d \, f \, m_3 C12 \, - \, 
 a \, d \, f \, m_1 \left[ C0 \, + \, C11 \right]  \end{eqnarray}
\bigskip

{\bf VERTEX PROTOTYPE 6}

\FIGURE[h]{
\centerline{\epsfig{file=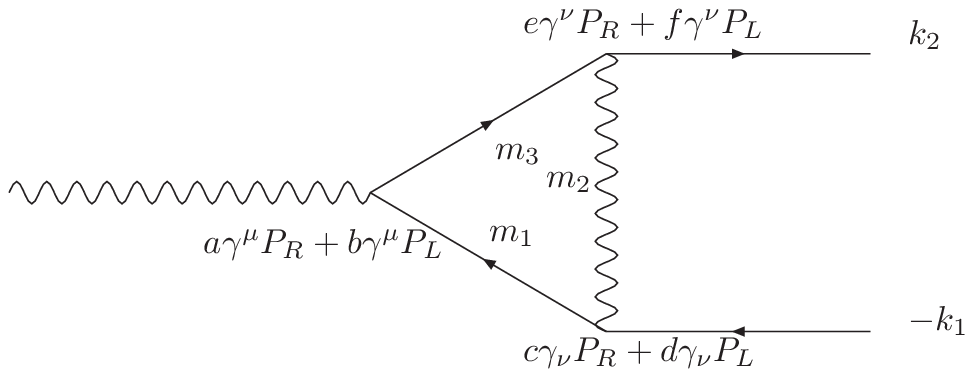}}
\caption{Vertex prototype graph 6.}
\label{figv6}
}

\begin{eqnarray} 16\pi^2 \, \Delta F_0^+ & = &  
  2 \,  a \, c \, e \,  \left[ 2 \, C24 \, - \, 1
 \, + \, \ma^2 \left(  C21 \, - \, C23 \right)
 \right. \nonumber \\ & & \left. \hspace*{1cm}
 \, + \, \mb^2 \left( C12 \, - \, C11 \, + \, C22 
\, - \, C23 \right) \, + \,  
s \left(C11 \, + C23 \right) \right] \nonumber \\
&-& 2 \, b \, d \, f \, \ma \mb \left[ C11 \, -  \, C12 \right]
  \, - \, 
2 \, b \, c \, e \, m_1 m_3  C0 
 \end{eqnarray}

\begin{eqnarray} 16\pi^2 \, \Delta F_0^- & = &  
  2 \,  b \, d \, f \,  \left[ 2 \, C24 \, - \, 1
 \, + \, \ma^2 \left(  C21 \, - \, C23 \right)
 \right. \nonumber \\ & & \left. \hspace*{1cm}
 \, + \, \mb^2 \left( C12 \, - \, C11 \, + \, C22 
\, - \, C23 \right) \, + \,  
s \left(C11 \, + C23 \right) \right] \nonumber \\
&-& 2 \, a \, c \, e \, \ma \mb \left[ C11 \, -  \, C12 \right]  \, - \, 
2 \, a \, d \, f \, m_1 m_3  C0 
 \end{eqnarray}

\begin{eqnarray} 16\pi^2 \, \Delta F_k^+ & = &  
  2 \,  b \, d \, f \, \ma \left[ C21 \, - \, C23 \right]
 \, + \, 2 \, a \, c \, e \, \mb \left[ C12 \, - \, C11 \, + \,
 C22 \, - \, C23 \right] \nonumber \\ &+&
2 \, b \, c \, f \, m_1 \left[C11 \, - \, C12 \right] \, + \, 
2 \, a \, c \, f \, m_3 \left[C11 \, - \, C12 \right]  \end{eqnarray}

\begin{eqnarray} 16\pi^2 \, \Delta F_k^- & = &  
  2 \,  a \, c \, e \, \ma \left[ C21 \, - \, C23 \right]
 \, + \, 2 \, b \, d \, f \, \mb \left[ C12 \, - \, C11 \, + \,
 C22 \, - \, C23 \right] \nonumber \\ &+&
2 \, a \, d \, e \, m_1 \left[C11 \, - \, C12 \right] \, + \, 
2 \, b \, d \, e \, m_3 \left[C11 \, - \, C12 \right]  \end{eqnarray}
\bigskip

{\bf VERTEX PROTOTYPE 7}

\FIGURE[h]{
\centerline{\epsfig{file=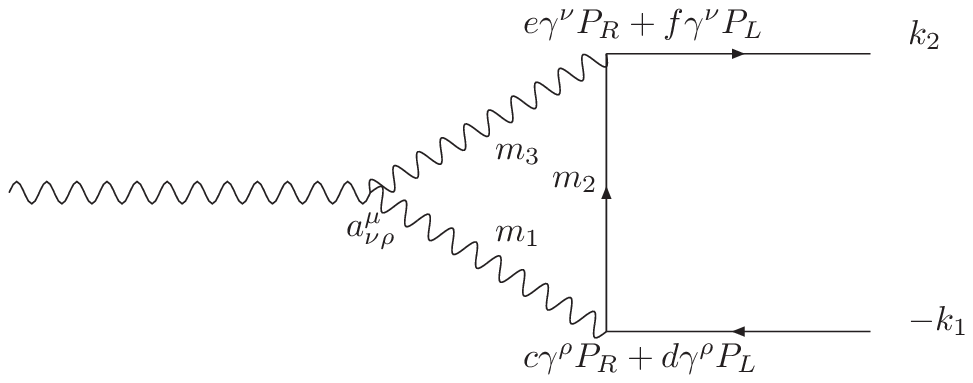}}
\caption{Vertex prototype graph 7.}
\label{figv7}
}

$a^\mu_{\nu\rho}$ is the usual triple gauge-boson vertex, multiplied
by a gauge coupling, $a$.

\begin{eqnarray} 16\pi^2 \, \Delta F_0^+ & = &  
    a \, c \, e \,  \left[ 4 \, C24 \, + \, 2 \, B0 \ \, - \, 1 
   \, + \, 2 \, m_1^2 C0
 \, + \, \ma^2 \left( 2  \, C0 \, + \, 
  4 \,  C11 \, - \, 3 \, C12 \right)
 \right. \nonumber \\ & & \left. \hspace*{1cm}
 \, + \, \mb^2 \left( C0 \,  + \, C11 \, - \, 2 \, C12  \right) \, + \,  
 s \left(3 \, C12 \, - \, C11 \, - \, C0  \right) \right] \nonumber \\
& & \hspace*{-2.3cm}  + \,
3 \, a \, d \, f \, \ma \mb \left[ C0 \, + \, C11 \, -  \, C12 \right]
\, - \, 3 \, a \, c \, f \, m_2 \mb C0 \, - \, 3 \, a  \, d \, e
 \,  m_2 \ma C0  \end{eqnarray}

\begin{eqnarray} 16\pi^2 \, \Delta F_0^- & = &  
    a \, d \, f \,  \left[ 4 \, C24 \, + \, 2 \, B0 \ \, - \, 1 
   \, + \, 2 \, m_1^2 C0
 \, + \, \ma^2 \left( 2  \, C0 \, + \, 
  4 \,  C11 \, - \, 3 \, C12 \right)
 \right. \nonumber \\ & & \left. \hspace*{-1cm}
 \, + \, \mb^2 \left( C0 \,  + \, C11 \, - \, 2 \, C12  \right) \, + \,  
 s \left(3 \, C12 \, - \, C11 \, - \, C0  \right) \right] \nonumber \\
& & \hspace*{-2.3cm} + \,
 3 \, a \, c \, e \, \ma \mb \left[ C0 \, + \, C11 \, -  \, C12 \right]
\, - \, 3 \, a \, d \, e \, m_2 \mb C0 \, - \, 3 \, a  \, c \, f
 \,  m_2 \ma C0  \end{eqnarray}

\begin{eqnarray} 16\pi^2 \, \Delta F_k^+ & = &  
a \, c \, e \, \mb \left[2 \, C0 \, + 2  \, C11 \, - \, C12
 - \, 2 \, C22 \, + 2 \, C23 \right]
 \nonumber \\ 
& & \hspace*{-3.3cm} + \,
  \, a \, d \,f \, \ma
\left[C0 \, - \, C11 \, - \, 2 \,  C21 \, + \, 2 \, C23 \right]
\, - \, 
3 \, a \, c \, f \, m_2 \left[ C0 \, - \, C11 \, + \, C12 \right] 
 \end{eqnarray}

\begin{eqnarray} 16\pi^2 \, \Delta F_k^- & = &  
a \, d \, f \, \mb \left[2 \, C0 \, + 2  \, C11 \, - \, C12
 - \, 2 \, C22 \, + 2 \, C23 \right]
 \nonumber \\ 
& & \hspace*{-3.3cm} + \,
 \, a \, c \,e \, \ma
\left[C0 \, - \, C11 \, - \, 2 \,  C21 \, + \, 2 \, C23 \right]
\, - \,  
3 \, a \, d \, e \, m_2 \left[ C0 \, - \, C11 \, + \, C12 \right] 
 \end{eqnarray}
\bigskip

{\bf VERTEX PROTOTYPE 8}

\FIGURE[h]{
\centerline{\epsfig{file=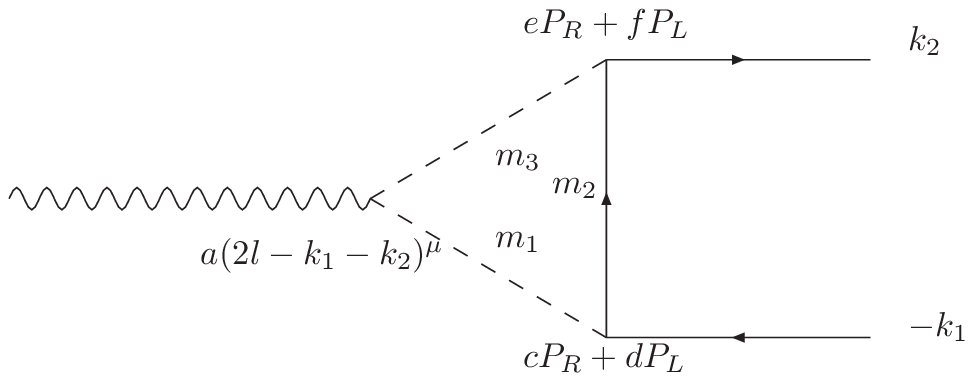}}
\caption{Vertex prototype graph 8.}
\label{figv8}
}

\beq 16\pi^2 \, \Delta F_0^+  =   
  - \, 2 \,  a \, c \, f \,  C24 \eeq

\beq 16\pi^2 \, \Delta F_0^-  =   
  - \, 2 \,  a \, d \, e \,  C24 \eeq

\begin{eqnarray} 16\pi^2 \, \Delta F_k^+ & = &  
a \, c \, f \, \mb \left[ C22 \, -   \, C23 \right]
 \, + \,   \, a \, d \, e  \, \ma
\left[ C11 \, - \, C12 \, + \,  C21 \, -  \, C23 \right]
\nonumber \\ &+&
 a \, c \, e \, m_2 \left[  C11 \, - \, C12 \right] 
 \end{eqnarray}

\begin{eqnarray} 16\pi^2 \, \Delta F_k^- & = &  
a \, d \, e \, \mb \left[ C22 \, -   \, C23 \right]
 \, + \,   \, a \, c \, f  \, \ma
\left[ C11 \, - \, C12 \, + \,  C21 \, -  \, C23 \right]
\nonumber \\ &+&
 a \, d \, f \, m_2 \left[  C11 \, - \, C12 \right] 
 \end{eqnarray}

\section*{Appendix D}
{\bf Prototype box graphs} \bigskip

\setcounter{equation}{0}
\renewcommand{\theequation}{D.\arabic{equation}}

In this Appendix we give the expressions for the prototype
box diagrams, with couplings indicated in the diagrams, in terms
of the Veltman-Passarino functions, $D0$, $D1(i), \ i=1\cdots 3$,
$D20$ and $D2(i,j), \ i,j=1\cdots 3$. These are defined as
\begin{eqnarray}
 - \, i \, \int \frac{d^4l}{\pi^2} \frac{1}
{(l^2-m_1^2)((l+v_1)^2-m_2^2) \, (l+v_1+v_2)^2-m_3^2) ((l-v_4)^2-m_4^2)}  & = &   
 \, D0  \nonumber \\
  - \, i \, \int \frac{d^4l}{\pi^2} \frac{l^\rho}
{(l^2-m_1^2)((l+v_1)^2-m_2^2) \, (l+v_1+v_2)^2-m_3^2) ((l-v_4)^2-m_4^2)}  & = &   
\sum_{i=1}^3  \, D1(i) v_i^\rho, \nonumber \\
  - \, i \, \int \frac{d^4l}{\pi^2} \frac{l^2}
{(l^2-m_1^2)((l+v_1)^2-m_2^2) \, (l+v_1+v_2)^2-m_3^2) ((l-v_4)^2-m_4^2)}  & = &  
 C0(1)+m_1^2  \, D0 \nonumber \\ 
  - \, i \,  \int \frac{d^4l}{\pi^2} \frac{l^\rho \, l^\sigma}
{(l^2-m_1^2)((l+v_1)^2-m_2^2) \, (l+v_1+v_2)^2-m_3^2) ((l-v_4)^2-m_4^2)}  & = &  
 g^{\rho\sigma}  \, D20 \nonumber \\  & & \hspace*{-2cm} 
 + \sum_{i,j=1}^3  
  \, D2(i,j) v_i^\rho v_j^\sigma \end{eqnarray}

For the uncrossed box prototypes 1, 2, 3a, 3b, 4a the vectors $v_i$ are
$$ v_1^\mu \ = \ p_1^\mu $$ 
$$ v_2^\mu \ = \ p_2^\mu $$ 
$$ v_3^\mu \ = \ -k_2^\mu $$
$$ v_4^\mu \ = \ -k_1^\mu $$
The arguments of the D-functions
are
 $$ Di(\alpha) = Di(\alpha)(s,t,0,0, m_{\chi_b}^2, m_{\chi_a}^2,m_1,m_2,m_3,m_4)$$
and $C0(1)$ means
 $$ C0(0,m_{\chi_b}^2,t,m_2^2,m_3^2,m_4^2), $$ 
whereas for the crossed box prototypes 4c,5a and 5b
$$ v_1^\mu=p_1^\mu$$
$$ v_2^\mu=-k_2^\mu$$
$$ v_3^\mu=p_2^\mu$$
$$ v_4^\mu=-k_1^\mu.$$
The arguments of the D-functions are 
 $$ Di(\alpha) = Di(\alpha)(u,t, 0, m_{\chi_b}^2,0, m_{\chi_a}^2,m_1,m_2,
m_3,m_4)$$
and $C0(1)$ means
 $$ C0(m_{\chi_b}^2,0,t,m_2^2,m_3^2,m_4^2) $$

The contributions, $\Delta{\mathcal Q}_{\alpha,i,j}$ to the
coefficient function  ${\mathcal Q}_{\alpha,i,j}$ from each prototype
box diagram are given below.
Diagrams involving internal gauge-bosons
are calculated in Feynman gauge.
\vspace*{1.5cm}

{\bf BOX PROTOTYPE 1}

\FIGURE[h]{
\centerline{\epsfig{file=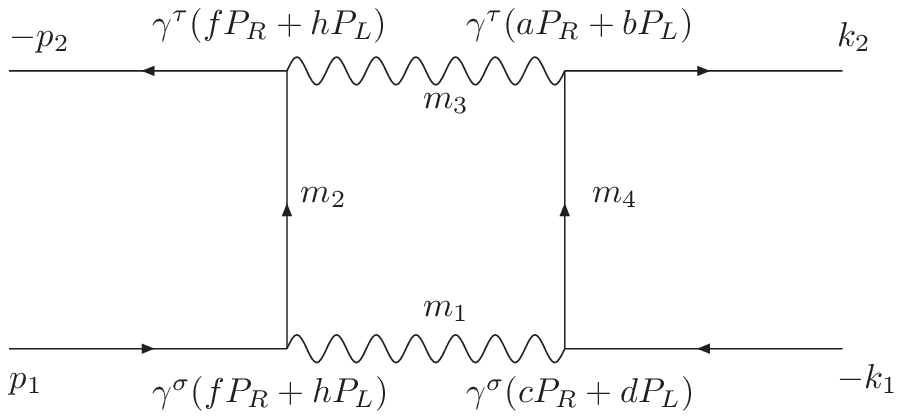}}
\caption{Box prototype graph 1.}
\label{figb1}
}

\begin{eqnarray} \frac{16\pi^2}{s}\Delta{\mathcal Q}_{R,1}& = & m_{\chi_a} 
\, b \, d \, f^2 \left[  \, D0 \, + \, D1(1) \, + \, D1(2)
  \, +  \, D1(3) \, + \, D2(3,1) \, + \, D2(3,2)\right] \nonumber \\ & - & 
 m_4 \, b \, c \, f^2 \left[  \, D0 \, + \, D1(1) \, + \, D1(2) \right] \end{eqnarray}

\begin{eqnarray} \frac{16\pi^2}{s}\Delta{\mathcal Q}_{R,2}& = &  \, - \, m_{\chi_b} 
\, b \, d \, f^2 \left[  \, D0 \, + \, D1(1)
  \, +  \, D1(2) \, - \, D1(3)
\right. \nonumber \\
& & \hspace*{1cm} \left. + \, D2(3,1) \, + \, D2(3,2) \, -2 \, D2(3,3) \right] \nonumber \\ &  \, + & 
 m_4 \, a \, d \, f^2 \left[  \, D0 \, + \, D1(1) \, + \, D1(2) \, -2 \, D1(3) \right]
 \end{eqnarray}

\begin{eqnarray} \frac{16\pi^2}{s}\Delta{\mathcal Q}_{R,3,1} &=&
 m_{\chi_a}  m_4 a \, d \, f^2 \left[  \, D0 \, + \, D1(1) \, + \, D1(2) \right] 
\nonumber \\ & \, -&
 2 m_{\chi_a}^2 a \, c \, f^2 \left[  \, D0 \, + \, D1(1) \, + \, D1(2) \, + \, D1(3) \right]
\nonumber \\ & \, -&
m_{\chi_b}  m_4 b \, c \, f^2 \left[  \, D0 \, + \, D1(1) \, + \, D1(2) \, -2  \, D1(3) \right] 
\nonumber \\ & -  &  
 4  \, m_1^2 a \, c \, f^2  \, D0 \nonumber \\ & & \hspace*{-2cm} + \,
a \, c \, f^2 \left[2 t  \left( D0 \,+ \, D1(1) \right)
  -4 \, C0(1)  \, -2 (s \, -u) ( \, D1(2) \, - \, D1(3))\right]
\end{eqnarray}

\beq\frac{16\pi^2}{s}\Delta{\mathcal Q}_{R,3,2}=0 \eeq

\begin{eqnarray} \frac{16\pi^2}{s}\Delta{\mathcal Q}_{R,4,1} &=&
 \, - \, m_{\chi_a} m_4 b \, c \, f^2 \left[ \, D0 \, + \, D1(1) \, + \, D1(2) \right]
\nonumber \\ & \, +&
 m_{\chi_b} m_4 a \, d \, f^2 \left[( \, D0 \, + \, D1(1) \, + \, D1(2) \, -2 \, D1(3) \right]
\nonumber \\ & \, -& 4 \, b \, d \, f^2  \, D20 \end{eqnarray}

\beq\frac{16\pi^2}{s}\Delta{\mathcal Q}_{R,4,2} = 
 \, - \,   b \, d \, f^2 \left[  \, D0 \, + \, D1(1) \, - \, D1(2) \, + \, D1(3) \, + \, D2(3,1) \, - \, D2(3,2) \right] \eeq
 
\beq\frac{16\pi^2}{s}\Delta{\mathcal Q}_{R,5,1} =  \, -\frac{1}{2} m_4 \left(a \, d  \, - b \, c\right) f^2
 \left[ \, D0 \, + \, D1(1) \, - \, D1(2) \right] \eeq 

\beq\frac{16\pi^2}{s}\Delta{\mathcal Q}_{R,5,2} =  \, - \frac{1}{4} m_4 \left(a \, d  \, + b \, c\right) f^2
 \left[ \, D0 \, + \, D1(1) \, - \, D1(2) \right] \eeq

\begin{eqnarray} \frac{16\pi^2}{s}\Delta{\mathcal Q}_{L,1}& = &  \, - \, m_{\chi_b} 
\, a \, c \, h^2 \left[  \, D0 \, + \, D1(1)
  \, +  \, D1(2) \, - \, D1(3) \right. \nonumber \\
& & \hspace*{1cm}  \left. + \, D2(3,1) \, + \, D2(3,2) \, -2 \, D2(3,3) \right] \nonumber \\ &  \, + & 
 m_4 \, b \, c \, h^2 \left[  \, D0 \, + \, D1(1) \, + \, D1(2) \, -2 \, D1(3) \right]
 \end{eqnarray}

\begin{eqnarray} \frac{16\pi^2}{s}\Delta{\mathcal Q}_{L,2}& = & m_{\chi_a} 
\, a \, c \, h^2 \left[  \, D0 \, + \, D1(1)  \, +  \, D1(2)
  \, +  \, D1(3) \, + \, D2(3,1) \, + \, D2(3,2)\right] \nonumber \\ &  \, - & 
 m_4 \, a \, d \, h^2 \left[  \, D0 \, + \, D1(1) \, + \, D1(2) \right] \end{eqnarray}

\begin{eqnarray} \frac{16\pi^2}{s}\Delta{\mathcal Q}_{L,3,1} &=&
 \, - \, m_{\chi_a} m_4 a \, d \, h^2 \left[ \, D0 \, + \, D1(1) \, + \, D1(2) \right]
\nonumber \\ & \, +&
 m_{\chi_b} m_4 b \, c \, h^2 \left[( \, D0 \, + \, D1(1) \, + \, D1(2) \, -2 \, D1(3) \right]
\nonumber \\ & \, -& 4 \, a \, c \, h^2  \, D20 \end{eqnarray}

\beq\frac{16\pi^2}{s}\Delta{\mathcal Q}_{L,3,2} = 
 \, - \,   a \, c \, h^2 \left[  \, D0 \, + \, D1(1) \, - \, D1(2) \, + \, D1(3) \, + \, D2(3,1) \, - \, D2(3,2) \right] \eeq

\begin{eqnarray} \frac{16\pi^2}{s}\Delta{\mathcal Q}_{L,4,1} &=&
 m_{\chi_a}  m_4 b \, c \, h^2 \left[  \, D0 \, + \, D1(1) \, + \, D1(2) \right] 
\nonumber \\ & \, -&
 2 m_{\chi_a}^2 b \, d \, h^2 \left[  \, D0 \, + \, D1(1) \, + \, D1(2) \, + \, D1(3) \right]
\nonumber \\ & \, -&
m_{\chi_b}  m_4 a \, d \, h^2 \left[  \, D0 \, + \, D1(1) \, + \, D1(2) \, -2  \, D1(3) \right] 
\nonumber \\ & \, -& 
 4  m_1^2 b \, d \, h^2  \, D0 \nonumber \\ & & \hspace*{-2.4cm} + \,
b \, d \, h^2 \left(2 t  \left( D0 \,+ \, D1(1) \right)
 \,  -4 \, C0(1) 
 \, - \, 2 (s \, -u) ( \, D1(2) \, - \, D1(3))\right]
\end{eqnarray}

\beq\frac{16\pi^2}{s}\Delta{\mathcal Q}_{L,4,2} = 0 \eeq

\beq\frac{16\pi^2}{s}\Delta{\mathcal Q}_{L,5,1} = \, \frac{1}{2} m_4 \left(a \, d  \, - b \, c\right) h^2
 \left[ \, D0 \, + \, D1(1) \, - \, D1(2) \right] \eeq 

\beq\frac{16\pi^2}{s}\Delta{\mathcal Q}_{L,5,2} = \frac{1}{4} m_4 \left(a \, d  \, + b \, c\right) h^2
 \left[ \, D0 \, + \, D1(1) \, - \, D1(2) \right] \eeq 
\bigskip

{\bf BOX PROTOTYPE 2}

\FIGURE[h]{
\centerline{\epsfig{file=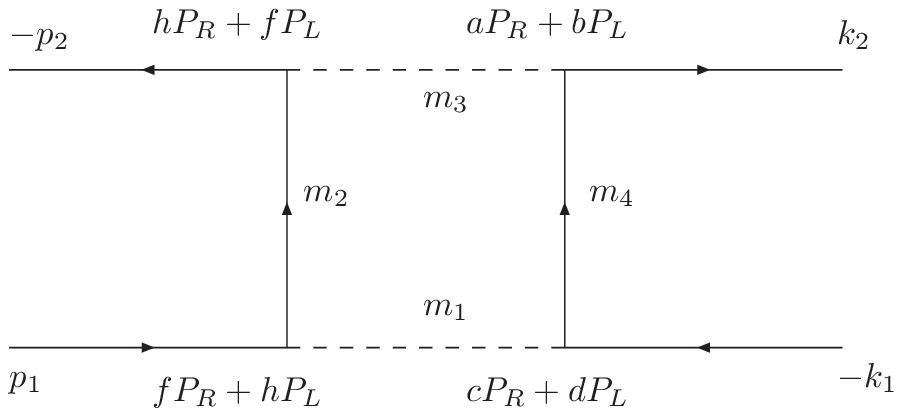}}
\caption{Box prototype graph 2.}
\label{figb2}
}

\begin{eqnarray} \frac{16\pi^2}{s}\Delta{\mathcal Q}_{R,1}& = & \frac{1}{4} m_{\chi_a} 
\, a \, d \, f^2 \left[ 2  \, D1(3) \, + \, D2(3,1)  \, + \, D2(3,2)\right] 
\nonumber \\ &  \, + & 
\,  \frac{1}{4} m_{\chi_b} 
\, b \, c \, f^2 \left[ 2  \, D2(3,3)- \, D2(3,1) - \, D2(3,2)\right] 
\nonumber \\ &  \, + & 
\frac{1}{2} \,  m_4 \, a \, c \, f^2
 \, D1(3)  \end{eqnarray}

\begin{eqnarray} \frac{16\pi^2}{s}\Delta{\mathcal Q}_{R,2}& = & \frac{1}{4} m_{\chi_a} 
\, b \, c \, f^2 \left[ 2  \, D1(3) \, + \, D2(3,1)  \, + \, D2(3,2)\right] 
\nonumber \\ &  \, + & 
\,  \frac{1}{4} m_{\chi_b} 
\, a \, d \, f^2 \left[ 2  \, D2(3,3) \, - \, D2(3,1)  \, - \, D2(3,2)\right] 
\nonumber \\ &  \, + & 
\frac{1}{2} \,  m_4 \, b \, d \, f^2
 \, D1(3)  \end{eqnarray}

\beq \frac{16\pi^2}{s}\Delta{\mathcal Q}_{R,3,1} =  \, - \, b \, c \, f^2  \, D20 \eeq

\beq\frac{16\pi^2}{s}\Delta{\mathcal Q}_{R,3,2}= \frac{1}{4} b \, c \, f^2
  \left[  \, D2(3,2) \, - \, D2(3,1)\right ] \eeq 

\beq \frac{16\pi^2}{s}\Delta{\mathcal Q}_{R,4,1} =  \, - \, a \, d \, f^2  \, D20 \eeq 

\beq\frac{16\pi^2}{s}\Delta{\mathcal Q}_{R,4,2}= \frac{1}{4} a \, d \, f^2
 \left[  \, D2(3,2) \, - \, D2(3,1) \right] \eeq 
 
\beq\frac{16\pi^2}{s}\Delta{\mathcal Q}_{R,5,1} = 0 \eeq

\beq\frac{16\pi^2}{s}\Delta{\mathcal Q}_{R,5,2} = 0 \eeq

\begin{eqnarray} \frac{16\pi^2}{s}\Delta{\mathcal Q}_{L,1}& = & \frac{1}{4} m_{\chi_a} 
\, a \, d \, h^2 \left[ 2  \, D1(3) \, + \, D2(3,1)  \, + \, D2(3,2)\right] 
\nonumber \\ &  \, + & 
\,  \frac{1}{4} m_{\chi_b} 
\, b \, c \, h^2 \left[ 2  \, D2(3,3) \, - \, D2(3,1)  \, - \, D2(3,2)\right] 
\nonumber \\ &  \, + & 
\frac{1}{2} \,  m_4 \, a \, c \, h^2
 \, D1(3)  \end{eqnarray}

\begin{eqnarray} \frac{16\pi^2}{s}\Delta{\mathcal Q}_{L,2}& = & \frac{1}{4} m_{\chi_a} 
\, b \, c \, h^2 \left[ 2  \, D1(3) \, + \, D2(3,1)  \, + \, D2(3,2)\right] 
\nonumber \\ &  \, + & 
\,  \frac{1}{4} m_{\chi_b} 
\, a \, d \, h^2 \left[ 2  \, D2(3,3) \, - \, D2(3,1)  \, - \, D2(3,2)\right] 
\nonumber \\ &  \, + & 
\frac{1}{2} \,  m_4 \, b \, d \, h^2
 \, D1(3)  \end{eqnarray}

\beq \frac{16\pi^2}{s}\Delta{\mathcal Q}_{L,3,1} =  \, - \, b \, c \, h^2  \, D20 \eeq

\beq\frac{16\pi^2}{s}\Delta{\mathcal Q}_{L,3,2}= \frac{1}{4} b \, c \, h^2
 \left[  \, D2(3,2) \, - \, D2(3,1)\right] \eeq 

\beq \frac{16\pi^2}{s}\Delta{\mathcal Q}_{L,4,1} =  \, - \, a \, d \, h^2  \, D20 \eeq 

\beq\frac{16\pi^2}{s}\Delta{\mathcal Q}_{L,4,2}= \frac{1}{4} a \, d \, h^2
 \left[  \, D2(3,2) \, - \, D2(3,1)\right] \eeq 
 
\beq\frac{16\pi^2}{s}\Delta{\mathcal Q}_{L,5,1} = 0 \eeq

\beq\frac{16\pi^2}{s}\Delta{\mathcal Q}_{L,5,2} = 0 \eeq
\bigskip

{\bf BOX PROTOTYPE 3a}

\FIGURE[h]{
\centerline{\epsfig{file=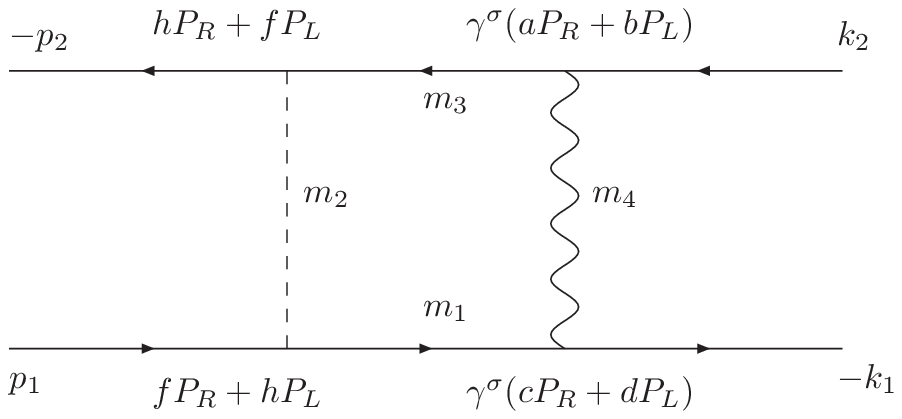}}
\caption{Box prototype graph 3a.}
\label{figb3a}
}

\begin{eqnarray} \frac{16\pi^2}{s}\Delta{\mathcal Q}_{R,1}& = &
 - \, \frac{1}{2} m_{\chi_a} 
\, b \, d \, f^2 \left[  \, 2  \, D1(3) \, + \, D2(3,1)  
\, + \, D2(3,2)\right] 
\nonumber \\
 &  \, + & 
 \,  m_3 \, a \, d \, f^2 \,  \, D1(3)  \end{eqnarray}

\begin{eqnarray} \frac{16\pi^2}{s}\Delta{\mathcal Q}_{R,2}& = & \frac{1}{2} m_{\chi_b} 
\, b \, d \, f^2 \left[  \, D2(3,1)  \, + \, D2(3,2) \, -2  \, D2(3,3)\right] 
\nonumber \\   &  \, + & 
  \,  m_1 \, b \, c \, f^2 \,  \, D1(3)  \end{eqnarray}

\beq \frac{16\pi^2}{s}\Delta{\mathcal Q}_{R,3,1} =  \, - m_{\chi_a} m_{\chi_b} \, b \, d \, f^2
  \, D1(3)  \, + m_1 m_3 \, a \, c \, f^2 \,  \, D0  \eeq

\beq\frac{16\pi^2}{s}\Delta{\mathcal Q}_{R,3,2}= 0 \eeq

\begin{eqnarray} \frac{16\pi^2}{s}\Delta{\mathcal Q}_{R,4,1} &=&
 \frac{1}{2} b \, d \, f^2 \left( m_{\chi_b}^2 \, -m_{\chi_a}^2 \right)  \, D1(3)
 \,  \, - \, m_1^2 b \, d \, f^2  \, D0 \nonumber \\ & \, +&
  \frac{1}{2} b \, d \, f^2 \left[ 4  \, D20 \, -2 \, C0(1) \, - (s \, -t \, +u)   \, D1(1)
  \right. \nonumber \\ & & \left.    \, - (s \, +t \, -u)
  \, D1(2) \, + (s \, -t \, -u)  \, D1(3) \right] \end{eqnarray}

\beq \frac{16\pi^2}{s}\Delta{\mathcal Q}_{R,4,2} = \frac{1}{2} \, b \, d \, f^2
 \left[   \, D1(1) \, - \, D1(2) \, + \, D2(3,1) \, - \, D2(3,2) \right] \eeq 

\beq\frac{16\pi^2}{s}\Delta{\mathcal Q}_{R,5,1}= \frac{1}{4} b \, d \, f^2 
\left(  m_{\chi_b} \, - m_{\chi_a} \right)
 \left[  \, D1(1) \, - \, D1(2) \right] \eeq 

\beq\frac{16\pi^2}{s}\Delta{\mathcal Q}_{R,5,2}= \frac{1}{8} b \, d \, f^2 
\left(  m_{\chi_b} \, + m_{\chi_a} \right)
 \left[  \, D1(1) \, - \, D1(2) \right] \eeq

\begin{eqnarray} \frac{16\pi^2}{s}\Delta{\mathcal Q}_{L,1}& = & \frac{1}{2} m_{\chi_b} 
\, a \, c \, h^2 \left[  \, D2(3,1)  \, + \, D2(3,2) \, -2 \, D2(3,3) \right] 
\nonumber \\ &  \, + & 
 \,  m_1 \, a \, d \, h^2  \, D1(3)  \end{eqnarray}

\begin{eqnarray} \frac{16\pi^2}{s}\Delta{\mathcal Q}_{L,2}& = &  \, - \frac{1}{2} m_{\chi_a} 
\, a \, c \, h^2 \left[2   \, D1(3)  \, + \, D2(3,1) \, + \, D2(3,2) \right] 
\nonumber \\ &  \, + & 
 \,  m_3 \, b \, c \, h^2  \, D1(3)  \end{eqnarray}

\begin{eqnarray} \frac{16\pi^2}{s}\Delta{\mathcal Q}_{L,3,1} &=&
 \frac{1}{2} a \, c \, h^2 \left( m_{\chi_b}^2 \, -m_{\chi_a}^2 \right)  \, D1(3)
 \,  \, - \, m_1^2 a \, c \, h^2  \, D0 \nonumber \\ & \, +&
  \frac{1}{2} a \, c \, h^2 \left[ 4  \, D20 \, -2 \, C0(1) \, - (s \, -t \, +u)   \, D1(1) \right. \nonumber \\ & & \left.
   \, - (s \, +t \, -u)   \, D1(2) \, + (s \, -t \, -u)  \, D1(3) \right] \end{eqnarray}

\beq\frac{16\pi^2}{s}\Delta{\mathcal Q}_{L,3,2}= \frac{1}{2} a \, c \, h^2
 \left[  \, D1(1) \, - \, D1(2) \, + \, D2(3,1) \, - \, D2(3,2)\right] \eeq 

\beq \frac{16\pi^2}{s}\Delta{\mathcal Q}_{L,4,1} =  \, -  m_{\chi_a} m_{\chi_b}  a \, c \, h^2  \, D1(3)
 \,  \, + \, m_1 m_3 \, b \ d \, h^2  \, D0  \eeq 

\beq\frac{16\pi^2}{s}\Delta{\mathcal Q}_{L,4,2}= 0 \eeq

\beq\frac{16\pi^2}{s}\Delta{\mathcal Q}_{L,5,1}=  \,  \frac{1}{4} a \, c \, h^2 
\left(  m_{\chi_b} \, - m_{\chi_a} \right)
 \left[  \, D1(1) \, - \, D1(2) \right] \eeq 

\beq\frac{16\pi^2}{s}\Delta{\mathcal Q}_{L,5,2}=  -\frac{1}{8} a \, c \, h^2 
 \, \left(  m_{\chi_b} \, + m_{\chi_a} \right)
 \left[  \, D1(1) \, - \, D1(2) \right] \eeq 
 \bigskip

{\bf BOX PROTOTYPE 3b}

\FIGURE[h]{
\centerline{\epsfig{file=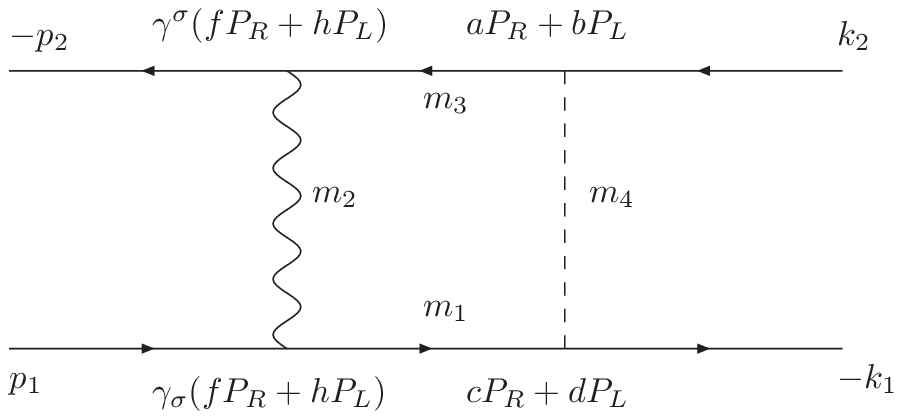}}
\caption{Box prototype graph 3b.}
\label{figb3b}
}

\begin{eqnarray} \frac{16\pi^2}{s}\Delta{\mathcal Q}_{R,1}& = &  \, -\frac{1}{2} m_{\chi_a} 
\, b \, c \, f^2 \left[ D2(3,1)  \, + \, D2(3,2)\right] 
\nonumber \\
 &  \, - & 
 \frac{1}{2} \,  m_3 \, a \, c \, f^2 \,\left[  \, D1(1) \, + \, D1(2) \right]
  \end{eqnarray}

\begin{eqnarray} \frac{16\pi^2}{s}\Delta{\mathcal Q}_{R,2}& = & \frac{1}{2} m_{\chi_b} 
\, b \, c \, f^2 \left[ 2 \, D1(3) \, + \, D2(3,1)  \, + \, D2(3,2) \, -2  \, D2(3,3)\right] 
\nonumber \\   &  \, + & 
  \,  \frac{1}{2} m_1 \, b \, d \, f^2 \, 
\left[ 2  \, D0 \, + \, D1(1) \, + \, D1(2) \, -2 \, D1(3) \right]
  \end{eqnarray}

\begin{eqnarray}
 \frac{16\pi^2}{s}\Delta{\mathcal Q}_{R,3,1}& =&  m_{\chi_a} m_{\chi_b} \, b \, c \, f^2
  \, D1(3)  \, + m_1 m_3 \, a \, d \, f^2 \,  \, D0  \nonumber \\ & \, +&
\frac{1}{2}  m_{\chi_a} m_1 \, b \, d \, f^2 \left[ 2 \, D0 \, + \, D1(1) \, + \, D1(2) \right] \nonumber \\ 
 & + & \frac{1}{2}  m_{\chi_b} m_3 \, a \, c \, f^2 \left[ 2 \, D1(3) \, - \, D1(1) \, - \, D1(2) \right]
\end{eqnarray}

\beq\frac{16\pi^2}{s}\Delta{\mathcal Q}_{R,3,2}= 0 \eeq

\begin{eqnarray} \frac{16\pi^2}{s}\Delta{\mathcal Q}_{R,4,1} &=&
  \frac{1}{2} m_{\chi_b} m_1  b \, d \, f^2 \left[ 2 \, D0 \, + \, D1(1) \, + \, D1(2) \, -2 \, D1(3)
    \right] \nonumber \\ & \, -&
  \frac{1}{2} m_{\chi_a} m_3  a \, c \, f^2 \left[  \, D1(1) \, + \, D1(2) \right]  \, - \, m_1^2 b \, c \, f^2  \, D0 
    \nonumber \\ & \, +&
  \frac{1}{2} b \, c \, f^2 \left[ 4  \, D20 \, -2 \, C0(1) \, - (s +t -u)   \, D1(1) \right. \nonumber \\ & & \left.
   \, - (s -t +u)   \, D1(2) \, + \, 2 \, \mb^2  \, D1(3) \right] \end{eqnarray}

\beq \frac{16\pi^2}{s}\Delta{\mathcal Q}_{R,4,2} = - \frac{1}{2} \, b \, c \, f^2
 \left[   \, D1(1) \, - \, D1(2) \, - \, D2(3,1) \, + \, D2(3,2) \right] \eeq 

\beq\frac{16\pi^2}{s}\Delta{\mathcal Q}_{R,5,1}= \frac{1}{4}
 f^2 
\left(  \left(  m_{\chi_a} \, - m_{\chi_b} \right) b \, c 
       - \, m_1 \, b \, d \, + \, m_3 \, a \, c \right) 
 \left[  \, D1(1) \, - \, D1(2) \right] \eeq 

\beq\frac{16\pi^2}{s}\Delta{\mathcal Q}_{R,5,2}= - \frac{1}{8}
 f^2 
\left(  \left(  m_{\chi_a} \, + m_{\chi_b} \right) b \, c 
      + \, m_1 \, b \, d \, + \, m_3 \, a \, c \right) 
 \left[  \, D1(1) \, - \, D1(2) \right] \eeq

\begin{eqnarray} \frac{16\pi^2}{s}\Delta{\mathcal Q}_{L,1}& = & \frac{1}{2} m_{\chi_b} 
\, a \, d \, h^2 \left[ 2 \, D1(3) \, +  \, D2(3,1)  \, + \, D2(3,2) 
\, -2 \, D2(3,3) \right] 
\nonumber \\ &  \, + & 
 \,  \frac{1}{2}m_1 \, a \, c\, h^2  \, 
\left[ 2 \, D0 \, + \, D1(1) \, + \, D1(2)\, - \, 2 \, 
D1(3) \right] \end{eqnarray}

\begin{eqnarray} \frac{16\pi^2}{s}\Delta{\mathcal Q}_{L,2}& = &  \, - \frac{1}{2} m_{\chi_a} 
\, a \, d \, h^2 \left[ \, D2(3,1) \, + \, D2(3,2) \right] 
\nonumber \\ &  - & 
 \frac{1}{2} \,  m_3 \, b \, d \, h^2 
\left[D1(1) \, + \, D1(2) \right]
 \end{eqnarray}

\begin{eqnarray} \frac{16\pi^2}{s}\Delta{\mathcal Q}_{L,3,1} &=&
  \frac{1}{2} m_{\chi_b} m_1  a \, c \, h^2 \left[ 2 \, D0 \, + \, D1(1) \, + \, D1(2) \, -2 \, D1(3)
    \right] \nonumber \\ & \, -&
  \frac{1}{2} m_{\chi_a} m_3  b \, d \, h^2 \left[  \, D1(1) \, + \, D1(2) \right] \, - \, m_1^2 a \, d \, h^2  \, D0
    \nonumber \\ & \, +&
  \frac{1}{2} a \, d \, h^2 \left[ 4  \, D20 \, -2 \, C0(1) \, - (s +t -u)   \, D1(1) \right. \nonumber \\   & & \left.
   \, - (s -t +u)   \, D1(2) \, + \, 2 \, \mb^2
  \, D1(3) \right] \end{eqnarray}

\beq\frac{16\pi^2}{s}\Delta{\mathcal Q}_{L,3,2}= -\frac{1}{2} a \, d \, h^2
 \left[  \, D1(1) \, - \, D1(2) \, - \, D2(3,1) \, + \, D2(3,2)\right] \eeq

\begin{eqnarray}
 \frac{16\pi^2}{s}\Delta{\mathcal Q}_{L,4,1} & =&  
 m_{\chi_a} m_{\chi_b} \, a \, d \, h^2
  \, D1(3)  \, +  
 m_1 m_3 \, b \, c \, h^2 \,  \, D0  \nonumber \\ & \, +&
\frac{1}{2}  m_{\chi_a} m_1 \, a \, c \, h^2 
\left[ 2 \, D0 \, + \, D1(1) \, + \, D1(2) \right] \nonumber \\ 
 & + & \frac{1}{2}  m_{\chi_b} m_3 \, b \, d \, h^2 
\left[ 2 \, D1(3) \, - \, D1(1) \, - \, D1(2) \right]
\end{eqnarray}

\beq\frac{16\pi^2}{s}\Delta{\mathcal Q}_{L,4,2}= 0 \eeq

\beq\frac{16\pi^2}{s}\Delta{\mathcal Q}_{L,5,1}= \frac{1}{4}
 h^2 
\left(  \left(  m_{\chi_a} \, - m_{\chi_b} \right) a \, d 
      - \, m_1 \, a \, c \, + \, m_3 \, b \, d \right) 
 \left[  \, D1(1) \, - \, D1(2) \right] \eeq 

\beq\frac{16\pi^2}{s}\Delta{\mathcal Q}_{L,5,2}=  \frac{1}{8}
 h^2 
\left(  \left(  m_{\chi_a} \, + m_{\chi_b} \right) a \, d 
      + \, m_1 \, a \, c \, + \, m_3 \, b \, d \right) 
 \left[  \, D1(1) \, - \, D1(2) \right] \eeq 
\bigskip

{\bf BOX PROTOTYPE 4a}

\FIGURE[h]{
\centerline{\epsfig{file=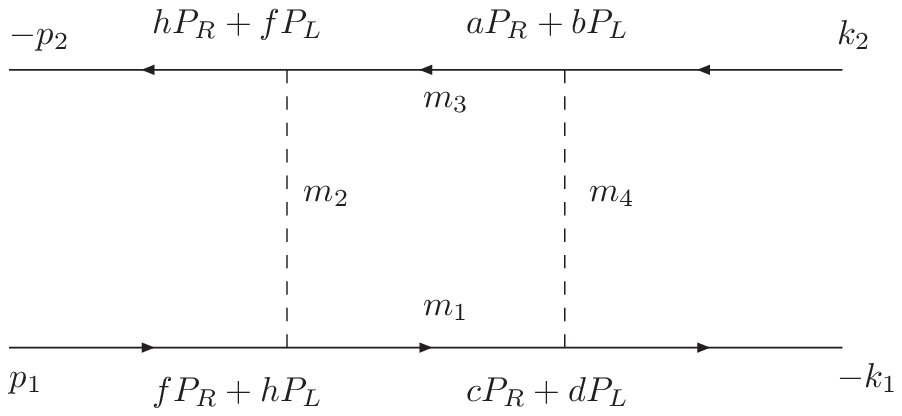}}
\caption{Box prototype graph 4a.}
\label{figb4a}
}

\begin{eqnarray} \frac{16\pi^2}{s}\Delta{\mathcal Q}_{R,1}& = &  \, \frac{1}{4} m_{\chi_b} 
\, a \, d \, f^2 \left[ 2 \, D1(3) \, + \, D2(3,1)
 \, + \, D2(3,2) \, - \, 2 \, D2(3,3)  \right] 
\nonumber \\
 &  \, + & 
 \frac{1}{4} \,  m_1 \, a \, c \, f^2 \,\left[2 \, D0 \, +
  \, D1(1) \, + \, D1(2) \, - \, 2 \, D1(3) \right]
  \end{eqnarray}

\begin{eqnarray} \frac{16\pi^2}{s}\Delta{\mathcal Q}_{R,2}& = & -\frac{1}{4} m_{\chi_a} 
\, a \, d \, f^2 \left[ D2(3,1)  \, + \, D2(3,2) \right] 
\nonumber \\   &  \, - & 
  \,  \frac{1}{4} m_3 \, b \, d \, f^2 \, 
\left[ D1(1) \, + \, D1(2)  \right]
  \end{eqnarray}

\begin{eqnarray} \frac{16\pi^2}{s}\Delta{\mathcal Q}_{R,3,1} &=&
  \frac{1}{4} m_{\chi_b} m_1  a \, c \, f^2 \left[
  2 \, D0 \, + \, D1(1) \, + \, D1(2) \, -2 \, D1(3)  \right]
 \nonumber \\ & \, -&
  \frac{1}{4 } m_{\chi_a} m_3  b \, d \, f^2 \left[
  \, D1(1) \, + \, D1(2) \right] \, - \frac{1}{2} \, m_1^2 a \, d \, f^2  \, D0
    \nonumber \\ &  +&
  \frac{1}{4} a \, d \, f^2 
   \left[ 4  \, D20 \, -2 \, C0(1) \, - (s +t -u)   \, D1(1)
 \right. \nonumber \\ & & \left.
    \, - (s -t +u)   \, D1(2) \, + \, 2 \, \mb^2 D1(3) \right] \end{eqnarray}

\beq\frac{16\pi^2}{s}\Delta{\mathcal Q}_{R,3,2}= - \frac{1}{4} 
 \, a \, d \, f^2 \left[ 
D1(1) \, - \, D1(2)\, - \, D2(3,1) \, + \, D2(3,2) \right]\eeq 

\begin{eqnarray}
 \frac{16\pi^2}{s}\Delta{\mathcal Q}_{R,4,1}& =&  \frac{1}{2} m_{\chi_a} m_{\chi_b} \, a \, d \, f^2
  \, D1(3)  \, + \, \frac{1}{2} m_1 m_3 \, b \, c \, f^2 \,  \, D0  \nonumber \\ & \, +&
\frac{1}{4}  m_{\chi_a} m_1 \, a \, c \, f^2 \left[
 2 \, D0 \, + \, D1(1) \, + \, D1(2) \right] \nonumber \\
 & + & \frac{1}{4}  m_{\chi_b} m_3 \, b \, d \, f^2 \left[ 2 \, D1(3) \, - \, D1(1) \, - \, D1(2) \right]
\end{eqnarray}

\beq \frac{16\pi^2}{s}\Delta{\mathcal Q}_{R,4,2} = 0 \eeq

\beq\frac{16\pi^2}{s}\Delta{\mathcal Q}_{R,5,1}= \frac{1}{8}  f^2 
\left(  \left(  m_{\chi_a} \, - m_{\chi_b} \right) a \, d 
       - \, m_1 \, a \, c \, + \, m_3 \, b \, d \right) 
 \left[  \, D1(1) \, - \, D1(2) \right] \eeq 

\beq\frac{16\pi^2}{s}\Delta{\mathcal Q}_{R,5,2}=  \frac{1}{16}
 f^2 
\left(  \left(  m_{\chi_a} \, + m_{\chi_b} \right) a \, d 
      + \, m_1 \, a \, c \, + \, m_3 \, b \, d \right) 
 \left[  \, D1(1) \, - \, D1(2) \right] \eeq

\begin{eqnarray} \frac{16\pi^2}{s}\Delta{\mathcal Q}_{L,1}& = & -\frac{1}{4} m_{\chi_a} 
\, b \, c \, h^2 \left[  D2(3,1)  \, + \, D2(3,2)  \right] 
\nonumber \\ &  - & 
 \,  \frac{1}{4} m_3 \, a \, c\, h^2  \, 
\left[   D1(1) \, + \, D1(2) \right] \end{eqnarray}

\begin{eqnarray} \frac{16\pi^2}{s}\Delta{\mathcal Q}_{L,2}& = &  \,  \frac{1}{4} m_{\chi_b} 
\, b \, c \, h^2 \left[2  \, D1(3) \, + \, D2(3,1) \, + \, D2(3,2) \, 
- \, 2 \, D2(3,3)  \right] 
\nonumber \\ &  + & 
 \frac{1}{4} \,  m_1 \, b \, d \, h^2 
\left[2 \, D0 \, + \, D1(1) \, + \, D1(2)\, - 2 \, D1(3) \right]
 \end{eqnarray}

\begin{eqnarray}
 \frac{16\pi^2}{s}\Delta{\mathcal Q}_{L,3,1}& =&  \frac{1}{2} m_{\chi_a} m_{\chi_b} \, b \, c \, h^2
  \, D1(3)  \, + \, \frac{1}{2} m_1 m_3 \, a \, d \, h^2 \,  \, D0  \nonumber \\ & \, +&
\frac{1}{4}  m_{\chi_a} m_1 \, b \, d \, h^2 \left[
 2 \, D0 \, + \, D1(1) \, + \, D1(2) \right] \nonumber \\
 & + & \frac{1}{4}  m_{\chi_b} m_3 \, a \, c \, h^2 \left[ 2 \, D1(3) \, - \, D1(1) \, - \, D1(2) \right]
\end{eqnarray}

\beq\frac{16\pi^2}{s}\Delta{\mathcal Q}_{L,3,2}= 0 \eeq

\begin{eqnarray} \frac{16\pi^2}{s}\Delta{\mathcal Q}_{L,4,1} &=&
  \frac{1}{4} m_{\chi_b} m_1  b \, d \, h^2 \left[ 2 \, D0 \, + \, D1(1) \, + \, D1(2) \, -2 \, D1(3)
    \right] \nonumber \\ & \, - &
  \frac{1}{4} m_{\chi_a} m_3  a \, c \, h^2 \left[  \, D1(1) \, + \, D1(2) \right]\, - \, \frac{1}{2} m_1^2 b \, c \, h^2  \, D0
    \nonumber \\ & \, +&
  \frac{1}{4} b \, c \, h^2 \left[ 4  \, D20 \, -2 \, C0(1) \, - (s +t -u)   \, D1(1) \right. \nonumber \\ & & \left.
   \, - (s -t +u)   \, D1(2) \, + \, 2 \, \mb^2 
 \, D1(3) \right] \end{eqnarray}

\beq\frac{16\pi^2}{s}\Delta{\mathcal Q}_{L,4,2}= - \frac{1}{4} \, 
b \, c \, h^2 \left[ D1(1) \, - \, D1(2) \, - \, D2(3,1)
 \, + \, D2(3,2) \right] \eeq

\beq\frac{16\pi^2}{s}\Delta{\mathcal Q}_{L,5,1}= \frac{1}{8}
 h^2 
\left(  \left(  m_{\chi_a} \, - m_{\chi_b} \right) b \, c 
      - \, m_1 \, b \, d \, + \, m_3 \, a \, c \right) 
 \left[  \, D1(1) \, - \, D1(2) \right] \eeq 

\beq\frac{16\pi^2}{s}\Delta{\mathcal Q}_{L,5,2}= - \frac{1}{16}
 h^2 
\left(  \left(  m_{\chi_a} \, + m_{\chi_b} \right) b \, c 
      + \, m_1 \, b \, d \, + \, m_3 \, a \, c \right) 
 \left[  \, D1(1) \, - \, D1(2) \right] \eeq 
\bigskip

{\bf BOX PROTOTYPE 4b}

\FIGURE[h]{
\centerline{\epsfig{file=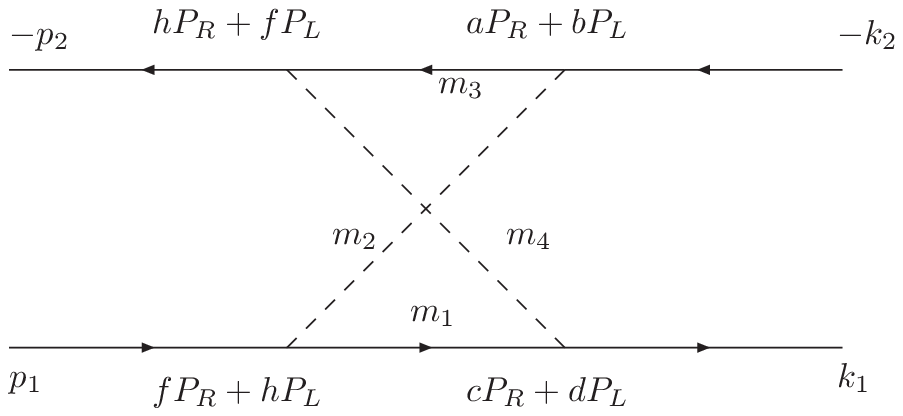}}
\caption{Box prototype graph 4b.}
\label{figb4b}
}

\begin{eqnarray} \frac{16\pi^2}{s}\Delta{\mathcal Q}_{R,1}& = &  \, \frac{1}{4} m_{\chi_b} 
\, a \, d \, f^2 \left[ D1(2) \, -\, D2(3,1) \, + \, 2 \,  D2(3,2) \,
 -  \, D2(3,3)  \right] 
\nonumber \\
 &  \, + & 
 \frac{1}{4} \,  m_1 \, a \, c \, f^2 \,\left[D0 \, - \,
   D1(1) \, + \, 2 \, D1(2) \,  -  \, D1(3) \right]
  \end{eqnarray}

\begin{eqnarray} \frac{16\pi^2}{s}\Delta{\mathcal Q}_{R,2}& = & \frac{1}{4} m_{\chi_a} 
\, a \, d \, f^2 \left[ D1(1) \, + \, D1(3)
  \, +   \, D2(3,1) \, +   \, D2(3,3) \right] 
\nonumber \\   &  \, - & 
  \,  \frac{1}{4} m_3 \, b \, d \, f^2 \, 
\left[D1(1) \, + \, D1(3)  \right]
  \end{eqnarray}

\begin{eqnarray} \frac{16\pi^2}{s}\Delta{\mathcal Q}_{R,3,1} &=&
 \frac{1}{8} a \, d \, f^2 \left( m_{\chi_a}^2 \, -m_{\chi_b}^2 \right)
  \left[D1(1) \, - \, D1(2) \, + \,  D1(3) \right]
\,  + \, \frac{1}{2} \, m_1^2 a \, d \, f^2  \, D0 \nonumber \\ &+&
  \frac{1}{4} m_{\chi_b} m_1  a \, c \, f^2 
\left[D0 \, - \,D1(1) \, + \, 2 \, D1(2) \, - \, D1(3) \right]
 \nonumber \\ &-&
  \frac{1}{4 } m_{\chi_a} m_3  b \, d \, f^2 
\left[  D1(1) \, + \, D1(3) \right]
    \nonumber \\ & \, - &
  \frac{1}{8} a \, d \, f^2 
   \left[ 8 \, D20 \, - \, 4 \, C0(1) + 
      (s +t -3 \, u)   \, D1(1) \right. \nonumber \\ & & \left.
   \, - (s +t +u)   \,( D1(2) \, + \, D1(3)) \right]
    \end{eqnarray}

\beq\frac{16\pi^2}{s}\Delta{\mathcal Q}_{R,3,2}= - \frac{1}{4} \, a \, d \, f^2 \left[ 
D1(1) \, + \, D2(3,1)\, - \, D2(3,3)  \right]\eeq 

\begin{eqnarray}
 \frac{16\pi^2}{s}\Delta{\mathcal Q}_{R,4,1}& =&  \frac{1}{4} m_{\chi_a} m_{\chi_b} \, a \, d \, f^2
\left[ D1(1) \, - \, D1(2) \, +  \, D1(3) \right]
  \, + \frac{1}{2} m_1 m_3 \, b \, c \, f^2 \,  \, D0  \nonumber \\ & \, -&
\frac{1}{4}  m_{\chi_a} m_1 \, a \, c \, f^2 \left[D0 \, + \,
  D1(1) \, + \, D1(3) \right] \nonumber \\
 & - & \frac{1}{4}  m_{\chi_b} m_3 \, b \, d \, f^2 
\left[ D1(1) \, - \, 2 \, D1(2) \, + \, D1(3) \right]
\end{eqnarray}

\beq \frac{16\pi^2}{s}\Delta{\mathcal Q}_{R,4,2} = 0 \eeq

\begin{eqnarray} 
\frac{16\pi^2}{s}\Delta{\mathcal Q}_{R,5,1} &=& -\frac{1}{8} a \, d \, f^2   m_{\chi_b}
 \left[D1(1) \, - \, D1(2) \right]
  +\frac{1}{8} a \, d \, f^2   m_{\chi_a} D1(3) \nonumber \\ & &
 \hspace*{-2.5cm} + \, 
\frac{1}{8} m_3 b \, d \, f^2 \left[D1(1) \, - \, D1(3) \right]
+\frac{1}{8} m_1 a \, c \, f^2 \left[D0 \, + \, D1(1) \, - \, D1(3) \right]
\end{eqnarray} 

\begin{eqnarray} 
\frac{16\pi^2}{s}\Delta{\mathcal Q}_{R,5,2} &=& \frac{1}{16} a \, d \, f^2   m_{\chi_b}
 \left[D1(1) \, - \, D1(2) \right]
  +\frac{1}{16} a \, d \, f^2   m_{\chi_a} D1(3) \nonumber \\ & &
 \hspace*{-3cm} + \, 
\frac{1}{16} m_3 b \, d \, f^2 \left[D1(1) \, - \, D1(3) \right]
- \frac{1}{16} m_1 a \, c \, f^2 \left[D0 \, + \, D1(1) \, - \, D1(3) \right]
\end{eqnarray}

\begin{eqnarray} \frac{16\pi^2}{s}\Delta{\mathcal Q}_{L,1}& = & \frac{1}{4} m_{\chi_a} 
\, b \, c\, h^2 \left[ D1(1) \, + \, D1(3)
  \, +   \, D2(3,1) \, +   \, D2(3,3) \right] 
\nonumber \\   &  \, - & 
  \,  \frac{1}{4} m_3 \, a \, c \, h^2 \, 
\left[D1(1) \, + \, D1(3)  \right]
  \end{eqnarray}

\begin{eqnarray} \frac{16\pi^2}{s}\Delta{\mathcal Q}_{L,2}& = &  \, \frac{1}{4} m_{\chi_b} 
\, b \, c \, h^2 \left[ D1(2) \, -\, D2(3,1) \, + \, 2 \,  D2(3,2) \,
 -  \, D2(3,3)  \right] 
\nonumber \\
 &  \, + & 
 \frac{1}{4} \,  m_1 \, b \, d \, h^2 \,\left[D0 \, - \,
   D1(1) \, + \, 2 \, D1(2) \,  - \, D1(3) \right]
  \end{eqnarray}

\begin{eqnarray}
 \frac{16\pi^2}{s}\Delta{\mathcal Q}_{L,3,1}& =&  \frac{1}{4} m_{\chi_a} m_{\chi_b} \, b \, c \, h^2
\left[ D1(1) \, - \, D1(2) \, +  \, D1(3) \right]
  \, + \frac{1}{2} m_1 m_3 \, a \, d \, h^2 \,  \, D0  \nonumber \\ & \, -&
\frac{1}{4}  m_{\chi_a} m_1 \, b \, d \, h^2 \left[D0 \, + \,
  D1(1) \, + \, D1(3) \right] \nonumber \\
 & - & \frac{1}{4}  m_{\chi_b} m_3 \, a \, c \, h^2 
\left[ D1(1) \, - \, 2 \, D1(2) \, + \, D1(3) \right]
\end{eqnarray}

\beq \frac{16\pi^2}{s}\Delta{\mathcal Q}_{L,3,2} = 0 \eeq

\begin{eqnarray} \frac{16\pi^2}{s}\Delta{\mathcal Q}_{L,4,1} &=&
 \frac{1}{8} b \, c \, h^2 \left( m_{\chi_a}^2 \, -m_{\chi_b}^2 \right)
  \left[D1(1) \, - \, D1(2) \, + \,  D1(3) \right]
\,  + \, \frac{1}{2} \, m_1^2 b \, c \, h^2  \, D0 \nonumber \\ &+&
  \frac{1}{4} m_{\chi_b} m_1  b \, d \, h^2 
\left[D0 \, - \,D1(1) \, + \, 2 \, D1(2) \, - \, D1(3) \right]
 \nonumber \\ &-&
  \frac{1}{4 } m_{\chi_a} m_3  a \, c \, h^2 
\left[  D1(1) \, + \, D1(3) \right]
    \nonumber \\ & \, - &
  \frac{1}{8} b \, c \, h^2 
   \left[ 8 \, D20 \, - \, 4 \, C0(1) + 
       (s +t -3 \, u)   \, D1(1) \right. \nonumber \\ & & \left.
   \, - (s +t +u)   \,( D1(2) \, + \, D1(3)) \right]
    \end{eqnarray}

\beq\frac{16\pi^2}{s}\Delta{\mathcal Q}_{L,4,2}= - \frac{1}{4} \, b \, c \, h^2 \left[ 
D1(1) \, + \, D2(3,1)\, - \, D2(3,3)  \right]\eeq

\begin{eqnarray} 
\frac{16\pi^2}{s}\Delta{\mathcal Q}_{L,5,1} &=& -\frac{1}{8} b \, c \, h^2   m_{\chi_b}
 \left[D1(1) \, - \, D1(2) \right]
  +\frac{1}{8} b \, c \, h^2   m_{\chi_a} D1(3) \nonumber \\ & &
 \hspace*{-2.5cm} + \,
\frac{1}{8} m_3 a \, c \, h^2 \left[D1(1) \, - \, D1(3) \right]
+\frac{1}{8} m_1 b \, d \, h^2 \left[D0 \, + \, D1(1) \, - \, D1(3) \right]
\end{eqnarray} 

\begin{eqnarray} 
\frac{16\pi^2}{s}\Delta{\mathcal Q}_{L,5,2} &=& -\frac{1}{16} b \, c \, h^2   m_{\chi_b}
 \left[D1(1) \, - \, D1(2) \right]
  -\frac{1}{16} b \, c \, h^2   m_{\chi_a} D1(3) \nonumber \\ & &
 \hspace*{-2.9 cm} - \,
\frac{1}{16} m_3 a \, c \, h^2 \left[D1(1) \, - \, D1(3) \right]
+ \frac{1}{16} m_1 b \, d \, h^2 \left[D0 \, + \, D1(1) \, - \, D1(3) \right]
\end{eqnarray} 
\bigskip

{\bf BOX PROTOTYPE 5a}

\FIGURE[h]{
\centerline{\epsfig{file=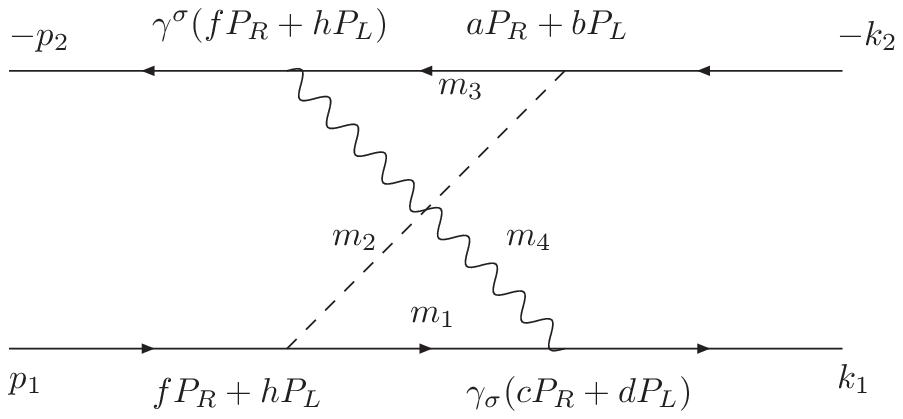}}
\caption{Box prototype graph 5a.}
\label{figb5a}
}

\begin{eqnarray} \frac{16\pi^2}{s}\Delta{\mathcal Q}_{R,1}& = & - \frac{1}{2} m_{\chi_a} 
\, b \, d\, f^2 \left[ D1(2) \, 
  \, +   \, D2(3,1) \, + \,  D2(3,3) \right] 
\nonumber \\   &  \, + & 
  \,  m_3 \, a \, d\, f^2 \, D1(2) 
   \end{eqnarray}

\begin{eqnarray} \frac{16\pi^2}{s}\Delta{\mathcal Q}_{R,2}& = &  \, -\frac{1}{2} m_{\chi_b} 
\, b \, d \, f^2 \left[ D1(2) \, -\, D2(3,1) \, + \, 2 \,  D2(3,2) \,
 -  \, D2(3,3)  \right] 
\nonumber \\
 &  \, + & 
 \frac{1}{2} \,  m_1 \, b \, c \, f^2 \,\left[D0 \, - \,
   D1(1) \, + \, 2 \, D1(2) \,  -   \, D1(3) \right]
  \end{eqnarray}

\beq \frac{16\pi^2}{s}\Delta{\mathcal Q}_{R,3,1} = -\frac{1}{2} m_{\chi_a} m_1 b \, c \, f^2
\left[ D0 \, + \, D1(1) \, + \, D1(3) \right] \, 
+ \, m_1m_3 a \, c \, f^2 D0 \eeq 

\beq \frac{16\pi^2}{s}\Delta{\mathcal Q}_{R,3,2} = 0 \eeq

\beq \frac{16\pi^2}{s}\Delta{\mathcal Q}_{R,4,1} = \frac{1}{2} m_{\chi_b} m_1 b \, c \, f^2
\left[ D0 \, - \, D1(1) \, + \, 2 \, D1(2) \, - \, D1(3) \right] \, 
+ \, 2 \, b \, d \, f^2  D20 \eeq

\beq\frac{16\pi^2}{s}\Delta{\mathcal Q}_{R,4,2}=  \frac{1}{2} \, b \, d \, f^2 \left[ 
D1(2) \, + \, D2(3,1)\, - \, D2(3,3)  \right]\eeq 

\beq\frac{16\pi^2}{s}\Delta{\mathcal Q}_{R,5,1}=\frac{1}{4} m_1 b \, c \, f^2
 \left[ D0 \, + \, D1(1) \, - \, D1(3) \right] \eeq

\beq\frac{16\pi^2}{s}\Delta{\mathcal Q}_{R,5,2}=\frac{1}{8} m_1 b \, c \, f^2
 \left[ D0 \, + \, D1(1) \, - \, D1(3) \right] \eeq

\begin{eqnarray} \frac{16\pi^2}{s}\Delta{\mathcal Q}_{L,1}& = & - \frac{1}{2} m_{\chi_b} 
\, a \, c \, h^2 \left[ D1(2) \, 
  \, -   \, D2(3,1) \, + 2 \,  \, D2(3,2) \,  - \,  D2(3,3) \right] 
\nonumber \\   &  \, + & 
  \,  \frac{1}{2} m_1 \, a \, d \, h^2 
\left[ D0 \, - \, D1(1) \, + \, 2 \, D1(2) \, - \, D1(3) \right]
  \end{eqnarray}

\beq \frac{16\pi^2}{s}\Delta{\mathcal Q}_{L,2} =   \, -\frac{1}{2} m_{\chi_a} 
\, a \, c \, h^2 \left[ D1(2) \, + \, D2(3,1) \, 
 + \, D2(3,3)  \right] \, + \, 
   m_3 \, b \, c \, h^2 \, D1(2) \eeq

\beq \frac{16\pi^2}{s}\Delta{\mathcal Q}_{L,3,1} = \frac{1}{2} m_{\chi_b} m_1 a \, d \, h^2
\left[ D0 \, - \, D1(1) \, + \, 2 \, D1(2) \, - \, D1(3) \right] \, 
+ \, 2 \, a \, c \, h^2 \, D20 \eeq

\beq\frac{16\pi^2}{s}\Delta{\mathcal Q}_{L,3,2}=  \frac{1}{2} \, a \, c \, h^2 \left[ 
D1(2) \, + \, D2(3,1)\, - \, D2(3,3)  \right]\eeq 

\beq \frac{16\pi^2}{s}\Delta{\mathcal Q}_{L,4,1} = -\frac{1}{2} m_{\chi_a} m_1 a \, d \, h^2
\left[ D0 \, + \, D1(1) \, + \, D1(3) \right] \, 
+ \, m_1 m_3 b \, d \, h^2 \, D0 \eeq

\beq \frac{16\pi^2}{s}\Delta{\mathcal Q}_{L,4,2} = 0 \eeq

\beq\frac{16\pi^2}{s}\Delta{\mathcal Q}_{L,5,1}=\frac{1}{4} m_1 a \, d \, h^2
 \left[ D0 \, + \, D1(1) \, - \, D1(3) \right] \eeq

\beq\frac{16\pi^2}{s}\Delta{\mathcal Q}_{L,5,2}=-\frac{1}{8} m_1 a \, d \, h^2
 \left[ D0 \, + \, D1(1) \, - \, D1(3) \right] \eeq
\bigskip

{\bf BOX PROTOTYPE 5b}

\FIGURE[h]{
\centerline{\epsfig{file=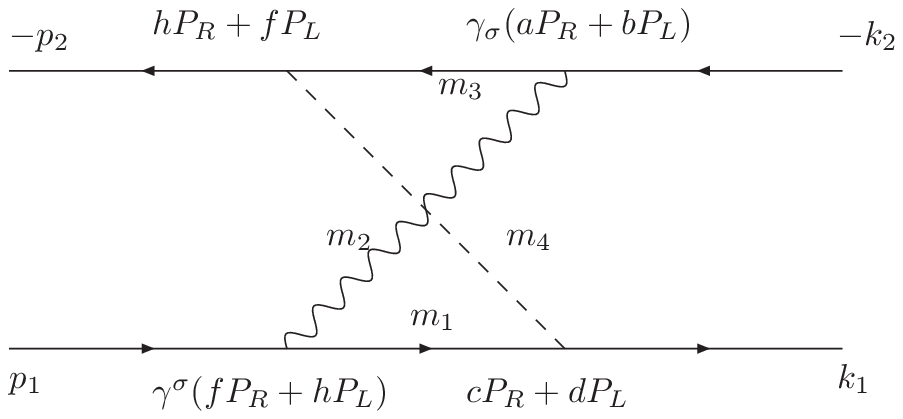}}
\caption{Box prototype graph 5b.}
\label{figb5b}
}

\begin{eqnarray} \frac{16\pi^2}{s}\Delta{\mathcal Q}_{R,1}& = & - \frac{1}{2} m_{\chi_a} 
\, b \, c \, f^2 \left[ D1(1) \, + \, D1(3)
  \, +   \, D2(3,1) \, + \,  D2(3,3) \right] 
\nonumber \\   &  \, - & 
  \,  \frac{1}{2} m_3 \, a \, c \, f^2 \, \left[D1(1) \, + \, D1(3) \right] 
  \end{eqnarray}

\begin{eqnarray} \frac{16\pi^2}{s}\Delta{\mathcal Q}_{R,2}& = &  \, \frac{1}{2} m_{\chi_b} 
\, b \, c \, f^2 \left[ D1(1) \, - \, 2 \, D1(2) \, + \, D1(3) \,
\right. \nonumber \\ & & \left. \hspace*{1cm}
  + \, D2(3,1) \, - \, 2 \,  D2(3,2) \, +  \, D2(3,3)  \right] 
\nonumber \\
 &  \, - & 
  m_1 \, b \, d \, f^2 
\left[D0 \, + \,  D1(2)  \right]
  \end{eqnarray}

\beq \frac{16\pi^2}{s}\Delta{\mathcal Q}_{R,3,1} = -\frac{1}{2} m_{\chi_b} m_3 a \, c \, f^2
\left[  D1(1) \, - \, 2 \, D1(2) \, + \, D1(3) \right] \, 
+ \, m_1m_3 a \, d \, f^2 D0 \eeq 

\beq \frac{16\pi^2}{s}\Delta{\mathcal Q}_{R,3,2} = 0 \eeq

\beq \frac{16\pi^2}{s}\Delta{\mathcal Q}_{R,4,1} = -\frac{1}{2} m_{\chi_a} m_3 a \, c \, f^2
\left[ D1(1) \, +  \, D1(3) \right] \, 
+ \, 2 \, b \, c \, f^2  D20 \eeq

\beq\frac{16\pi^2}{s}\Delta{\mathcal Q}_{R,4,2}=  \frac{1}{2} \, b \, c \, f^2 \left[ 
D1(1) \, - \, D1(3)  \, + \, D2(3,1)\, - \, D2(3,3)  \right]\eeq 

\beq\frac{16\pi^2}{s}\Delta{\mathcal Q}_{R,5,1}=\frac{1}{4} m_3 a\, c \, f^2
 \left[  D1(1) \, - \, D1(3) \right] \eeq

\beq\frac{16\pi^2}{s}\Delta{\mathcal Q}_{R,5,2}=- \frac{1}{8} m_3 a \, c \, f^2
 \left[ D1(1) \, - \, D1(3) \right] \eeq

\begin{eqnarray} \frac{16\pi^2}{s}\Delta{\mathcal Q}_{L,1}& = &  \frac{1}{2} m_{\chi_b} 
\, a \, d \, h^2 \left[ D1(1) \, - \, 2 \, D1(2) \, + \, D1(3)  
  \,  +  \, D2(3,1) \, - 2 \,  \, D2(3,2) \,  +\,  D2(3,3) \right] 
\nonumber \\   &  \, - & 
  \,   m_1 \, a \, c \, h^2 
\left[ D0 \, + \, D1(2)  \right]
  \end{eqnarray}

\begin{eqnarray}  \frac{16\pi^2}{s}\Delta{\mathcal Q}_{L,2} &=&   \, -\frac{1}{2} m_{\chi_a} 
\, a \, d \, h^2 \left[D1(1) \, + \,  D1(3) \, + \, D2(3,1) \, 
  + \, D2(3,3)  \right] 
\nonumber \\ 
& - & 
 \frac{1}{2} \,  m_3 \, b \, d \, h^2 \, 
\left[ D1(1) \, + \, D1(3) \right] \end{eqnarray}

\beq \frac{16\pi^2}{s}\Delta{\mathcal Q}_{L,3,1} = -\frac{1}{2} m_{\chi_a} m_3 b \, d \, h^2
\left[  D1(1) \, +  D1(3) \right] \, 
+ \, 2 \, a \, d \, h^2 \, D20 \eeq

\beq\frac{16\pi^2}{s}\Delta{\mathcal Q}_{L,3,2}=  \frac{1}{2} \, a \, d \, h^2 \left[ 
D1(1) \, - \, D1(3)  \, + \, D2(3,1)\, - \, D2(3,3)  \right]\eeq 

\beq \frac{16\pi^2}{s}\Delta{\mathcal Q}_{L,4,1} = -\frac{1}{2} m_{\chi_b} m_3 b \, d \, h^2
\left[ D1(1) \, - \, 2 \, D1(2)  \, + \, D1(3) \right] \, 
+ \, m_1 m_3 b \, c \, h^2 \, D0 \eeq

\beq \frac{16\pi^2}{s}\Delta{\mathcal Q}_{L,4,2} = 0 \eeq

\beq\frac{16\pi^2}{s}\Delta{\mathcal Q}_{L,5,1}=\frac{1}{4} m_3 b \, d \, h^2
 \left[ D1(1) \, - \, D1(3) \right] \eeq

\beq\frac{16\pi^2}{s}\Delta{\mathcal Q} _{L,5,2}=\frac{1}{8} m_3 b \, d \, h^2
 \left[ D1(1) \, - \, D1(3) \right] \eeq

\bigskip
All other box diagram prototypes that give contributions to these amplitudes
can be obtained from the above box  prototypes by means of crossing
relations (e.g. $u \leftrightarrow t$, and $\ma \leftrightarrow \mb$).

\end{document}